\documentclass[12pt]{iopart}

\newcommand{\TT}{ \frac{T_{bg} } { T }  }
\usepackage{iopams,graphicx,verbatim}  
\begin{document}



\title[Uncertainty Intervals for Weak Poisson Signals]{
Frequentist Coverage Properties of Uncertainty Intervals for
Weak Poisson Signals 
in the Presence of Background
}
\author{K. J. Coakley}
\address{National Institute of Standards and Technology, Boulder, CO}
\ead{kevin.coakley@nist.gov}
\author{J. D. Splett}
\address{National Institute of Standards and Technology, Boulder, CO}
\author{D. S. Simons}
\address{National Institute of Standards and Technology, Gaithersburg, MD}
\date{\today}




\begin{abstract}
We construct
uncertainty intervals
for weak Poisson signals
in the presence of
background.
We consider the case where
a primary experiment
yields a realization of the signal plus background,
and a second experiment yields a realization of
the background.
The data acquisitions times,
for the background-only experiment, $T_{bg}$,
and the primary experiment, $T$,
are selected so that their ratio, $\TT$, varies from
1 to 25.
The upper choice of 25
is motivated by an experimental study
at the National Institute of Standards and Technology (NIST).
The expected number of background counts in the primary experiment varies from 0.2 to 2.
We construct 90 and 95 percent confidence intervals
based on a propagation-of-errors method
as well as two implementations of a Neyman procedure
where acceptance regions are constructed based on
a likelihood-ratio criterion
that automatically determines whether the
resulting confidence interval is one-sided or
two-sided.
In one of the implementations of the Neyman procedure due to
Feldman and Cousins, uncertainty in the expected background
contribution is neglected. In the other implementation,
we account for random uncertainty in the estimated expected 
background with a parametric bootstrap implementation
of a method due to Conrad.
We also construct
minimum length Bayesian credibility intervals.
For each method,
we test
for the presence of a signal
based on the value of the lower endpoint of the uncertainty interval.
In general, the propagation-of-errors method performs
the worst compared to the other methods according to 
frequentist coverage 
and detection probability criteria, and sometimes
produces nonsensical intervals where both endpoints
are negative.
The Neyman procedures generally yield
intervals with
better frequentist coverage properties compared to
the Bayesian method except for some cases where
$\TT = 1$.
In general, the Bayesian method yields intervals
with lower detection probabilities
compared to
Neyman procedures.
One of main conclusions is that
when
$\TT $ is 5 or more and
the expected background is 2 or less,
the FC method
outperforms the other methods considered.
For
$\frac{T_{bg}}{T} =1,2$
we observe that
the Neyman procedure methods yield 
false detection probabilities for the case of no signal
that are higher than expected given the
nominal frequentist coverage of the interval.
In contrast, for 
$\frac{T_{bg}}{T} =1,2$, the false detection probability of
the Bayesian method is less than expected according
to the nominal frequentist coverage.
\pacs{02.50.-r,07.90.+c,07.05.Kf,07.81.+a,14.60.Lm,29.85.-c,29.85.Fy,95.35.+d}
\end{abstract}
\small
\noindent
keywords:
astroparticle and particle physics,
background contamination,
data and error analysis,
isotopic ratios,
low level radiation detection,
metrology and the theory of measurement,
Poisson processes,
signal detection,
uncertainty intervals.
\noindent
Contributions by staff of the National Institute of Standards
and Technology, an agency of the US government, are not subject
to copyright.
\normalsize


\maketitle


\section{Introduction}
We consider
experiments where instruments
yield 
count data
that can
be  modeled as
realizations of a Poisson process with
expected value
$\mu_S + \mu_B$ where
$\mu_S$ is the expected contribution due to a signal of interest,
and $\mu_B$ is the expected contribution of a background process.
That is,
\begin{eqnarray}
n_{obs} \sim Poi ( \mu_S  + \mu_B).
\end{eqnarray}
Given the measured value $n_{obs}$ and
an estimate of $\mu_B$ from an independent background-only
experiment,
we construct
uncertainty intervals
(confidence intervals and
Bayesian credibility intervals)
for $\mu_S$. 
The statistical problem we study occurs in
a variety of application areas including:
particle and astroparticle physics 
[1-9],
isotopic 
ratio analysis (when the major isotope is large enough
so
that most of the variability in the ratio is due to the minor isotope)
[10,11],
detection of 
low-level radiation 
[12-15], and aerosol science and technology [16].

Here, we focus on the case
where the signal is weak and
consider the case where the ratio of the data acquisition time for the
background-only measurement $T_{bg}$  and the 
data acquisition time 
for the primary experiment $T$ varies from 1 to 25.
This upper value of $\TT = 25$  
was motivated by an experimental study at
the National Institute of Standards and Technology (NIST), as were
the values of $\mu_{S}$ and $\mu_{B}$  that we consider.
For such cases,
we demonstrate
that the standard
propagation-of-errors (POE) method 
yields confidence intervals
with poor coverage properties. Sometimes the
POE method produces
intervals
where
the upper and/or lower endpoints are negative.
As an aside,
for the special case where
$\TT = 1$,
one can
construct a confidence interval for $\mu_S$
based on a Bessel function approach
that has better coverage properties than
does the POE method [16].
However, for the general case where
$\TT \ne 1$,
this method
is not applicable.
Hence, we do not include the method described in [16]
in our study.

In addition to the POE method,
we study the relative performance of
three other methods for constructing 
uncertainty intervals.
The first method [17]
is an implementation of 
a frequentist Neyman procedure
[18]
developed by Feldman and Cousins.
In this method, which we refer to as the FC method,
$\mu_B$ is assumed to be known.
For each of many discrete values of
$\mu_S$, acceptance regions are
constructed
based on a likelihood-ratio criterion.
Given the intersection of the actual measured value 
with these regions, one constructs a confidence interval for
$\mu_S$.
In our studies,
we estimate $\mu_B$ 
from background-only experiments.
In [19],
the FC method was extended to account for 
systematic uncertainties in $\mu_B$.
We denote this method as the randomized Feldman Cousins (RFC) method
because
$\mu_B$ is treated as a random nuisance parameter.
In this work, we implement a version of the RFC method where uncertainty in
$\mu_B$ is
due to Poisson counting statistics variation in
a background-only experiment that gives a direct measurement of $\mu_B$.
In the RFC method,
we simulate realizations of the nuisance parameter $\mu_B$
with a parametric bootstrap method [20].
In both
the FC and the RFC
methods,
the upper and lower endpoints are determined automatically. 	

We also determine the 
posterior probability density
function (posterior pdf) for $\mu_S$ with a Bayesian method  [21,22]
following 
Loredo's treatment of the same problem in
[23].
Loredo did not discuss how to 
select the endpoints of
the credibility interval. 
Here,
given that the integrated posterior pdf has
a particular value (equal to the nominal
frequentist coverage probability),
we determine the endpoints by
minimizing the length of the credibility
interval. 
As an aside for the special case where
$\mu_B$ is known,
Roe and Woodroofe [24]  
determined minimum length Bayesian
credibility intervals assuming a uniform prior
for $\mu_S$ and studied the frequentist
coverage properties of their intervals.

Bayesian credibility intervals and frequentist
confidence intervals are
conceptually different.
To illustrate, consider a one-dimensional parameter estimation
problem.
Frequentist confidence intervals are constructed so that,
ideally, the true value of the parameter
falls within the confidence interval
determined from any independent realization
of data with some desired coverage probability.
In contrast, Bayesian credibility intervals 
are
constructed by modeling the
parameter of interest
as a random variable.
Given a
prior probability model for the parameter of interest
and 
a likelihood model
for the data given the parameter, Bayes theorem
yields
the conditional probability
density function of the parameter given the observed data.
Based on this conditional pdf (called the posterior pdf),
one constructs credibility intervals.
By design, Bayesian credibility intervals are not
constructed with frequentist coverage
in mind.
Although the
conceptual
foundations of frequentist and Bayesian inference are different,
frequentist coverage is widely 
accepted as an empirical measure of
the performance of not only frequentist confidence intervals
but
Bayesian credibility intervals as well [22,25,26].
In a highly regarded textbook on
Bayesian data analysis, Gelman, Carlin, Stern and Rubin
remark (page 111 of [22])
\begin{quote}
Just as the Bayesian paradigm can be seen to 
justify simple `classical' techniques, the 
methods of frequentist statistics provide a useful
approach for evaluating the properties of 
Bayesian inferences- their operating characteristics--
when these are regarded as embedded in a sequence of
repeated samples.
\end{quote}

In this frequentist coverage study, we
simulate realizations of data given $\mu_S$ and $\mu_B$,
and
quantify the probability 
that $\mu_S$ falls in the interval determined from
the simulated data.
In frequentist statistics, the relationship between
a confidence interval and a hypothesis test is well 
known.
We exploit this relationship and test the null hypothesis that
$\mu_S= 0$
against the alternative hypothesis $\mu_S > 0$,
based on the value of the lower endpoint of the
uncertainty interval.
We reject the null hypothesis if the lower endpoint
is greater than 0.
Thus,
the probability  
that the lower endpoint of an interval
is greater than 0 is a signal detection probability.
As a caveat, we do not claim that this
procedure is the most powerful test of our hypothesis.

In Section 2, we define our measurement model
and describe how we determine
uncertainty intervals using each of the four methods.
In this study,
the background  parameter $\mu_B$ ranges
from 0.2 to 2 and
the signal parameter $\mu_S$ ranges from
0 to 20.
In Section 3, we study the coverage properties of
uncertainty intervals for a variety of cases.
We also determine detection probabilities
for a signal of interest.
In general, the propagation-of-errors method performs
the worst compared to the other methods according to 
frequentist coverage 
and detection probability criteria.
Further, the 
propagation-of-errors method sometimes 
produces nonsensical intervals where both endpoints
are negative.
The Neyman procedures generally yield
intervals with
better frequentist coverage properties compared to
the Bayesian method except for the case where
$\TT = 1$
and there are 1 or more expected background
counts in the primary experiment.
In general, the Bayesian method yields intervals
with lower detection probabilities
compared to
Neyman procedures.
When
$\TT $ is 5 or more,
the FC method
yields intervals with the highest detection probabilities
and best coverage properties in general.
However, for
$ \frac{T_{bg}}{T} =1,2$
both the Neyman procedure methods yield 
false detection probabilities for the case of no signal
that are higher than expected given the
nominal frequentist coverage of the interval.
In contrast, for 
$\frac{T_{bg}}{T} =1,2$, the false detection probability of
the Bayesian method is less than expected according
to the nominal frequentist coverage.

\section{Measurement Model and Uncertainty Intervals }

In our simulation study,
we consider an experiment where a realization of
the signal of interest
plus background, $n_{obs}$ is observed during
a time interval
$T$.
In a separate experiment of duration $T_{bg}$, where
$ \TT $ varies from 1 to 25,
we measure a realization of background 
$n_{bg}$.
We denote the data as $d= (n_{obs},n_{bg})$.
The expected values of $n_{obs}$ and $n_{bg}$ are
$\mu_S + \mu_B$ and $ \mu_B \TT$,
respectively, where
$\mu_S$ is the expected contribution from the signal of interest, 
and $\mu_B$ is the expected contribution from the background.
We model measurements  of $n_{obs}$ and $n_{bg}$
as independent Poisson random variables.
Hence, the likelihood function of the data is $P(d | \mu_S, \mu_B)$, where
\begin{eqnarray}
P(d|\mu_S,\mu_B) = (\mu_S + \mu_B ) ^{n_{obs}} \frac{ \exp[ - ( \mu_S + \mu_B )  ] } { n_{obs} !}
\times
\end{eqnarray}
\begin{eqnarray*}
( \mu_B \TT ) ^{n_{bg}} \frac{ \exp[ - ( \mu_B \TT)  ] } { n_{bg} !}.
\end{eqnarray*}

\subsection{Feldman Cousins Method}

In the FC method,
one determines
confidence intervals
with a Neyman procedure assuming 
exact knowledge of
$\mu_B$.
In our study,
we set $\mu_B$ to an empirical estimate $\hat{\mu}_B$,
where
\begin{eqnarray}
\hat{\mu}_B = \frac{ T } { T_{bg} } n_{bg}.
\end{eqnarray}
Hence,
the variance 
of
$\hat{\mu}_B$ is
\begin{eqnarray}
VAR ( \hat{\mu}_B ) = (\frac{ T} { T_{bg} } )  \mu_B,
\end{eqnarray}
and the standard deviation of
$\hat{\mu}_B$ is
\begin{eqnarray}
\sigma ( \hat{\mu}_B ) =\sqrt{ \frac{ T} { T_{bg} } }  \sqrt{\mu_B}.
\end{eqnarray}
Thus, the fractional uncertainty of the estimate of
$\mu_B$
is
\begin{eqnarray}
\frac{
\sigma ( \hat{\mu}_B )
}
{\mu_B}
=\sqrt{ \frac{ T} { T_{bg} } }  \frac{1} { \sqrt{\mu_B}}.
\end{eqnarray}

In Figure 1, we plot probability density functions for
the estimated background 
when
$\frac{ T_{bg}} { T }  = $ 25 
for $\mu_B=$ 0.2, 1, and 2.

The FC method [17] 
produces a confidence interval for $\mu_S$ under the assumption that the assumed background ($\hat{\mu}_B$ in our case)
equals the true background $\mu_B$. 
For
various values of
$\mu_S$,
we construct an acceptance region in
$n$ space.
For each integer value of $n$,
we compute the conditional probability 
$P(n | \mu_S, \hat{\mu}_B)$ and
$P(n| \mu_S, \hat{\mu}_{best})$,
where $\hat{\mu}_{best}= max(0,n-\hat{\mu}_B)$
and
\begin{eqnarray}
P(n|\mu_S, \hat{\mu}_B) = (\mu_S + \hat{\mu}_B ) ^n \frac{ \exp[ - ( \mu_S + \hat{\mu}_B )  ] } { n !}.
\end{eqnarray}

From these, we form the ratio $R$, where
\begin{eqnarray}
R = \frac{ P(n | \mu_S, \hat{\mu}_B) } { P ( n | \mu_S, \hat{\mu}_{best} ) }.
\end{eqnarray}
We include values of $n$
in the acceptance region with 
the largest values of $R$.
For construction of a 100 $\times ~ p$  $\%$
confidence interval, we add values until
the sum of the $P(n | \mu_S, \hat{\mu}_B)$ terms is 
$p$ or greater.
The lower 
and upper
endpoints of the
confidence interval for $\mu_S$
are the 
minimum and maximum 
values of $\mu_S$ that yield acceptance regions that
include the observed value $n_{obs}$.
For fixed $n_{obs}$,
due to the discreteness of $n$,
the upper endpoint of the interval
is not always a decreasing function of $\mu_B$.
In [17], Feldman and Cousins lengthened their intervals
so that the upper interval was a non-decreasing function of
$\mu_B$.
In this work, we do not adjust
our
intervals.

\subsection{Extension of Feldman Cousins: Uncertain Background}
In the RFC method,
the value of
$\mu_B$ is a random nuisance parameter.
In our analysis, we assume that uncertainty in $\mu_B$ is due to 
random variation alone, i.e., counting statistics. If there were systematic error, 
it could be incorporated into the analysis. 
However, we do not do this.

The procedure to construct a confidence interval is very similar
to the FC method.
For each value of $\mu_S$,
we compute an acceptance region 
like before, but
we replace
$P(n | \mu_S, \hat{\mu}_B) $
and
$P ( n | \mu_S, \hat{\mu}_{best} ) $
with an estimate of
their expected values when one accounts for uncertainty in
$\hat{\mu}_B$.

One way to do this would be to simulate
realizations of
$\hat{\mu}_B$
with a parametric bootstrap [20]
method and determine the mean value of
$P(n | \mu_S, \hat{\mu}_B) $
and
$P ( n | \mu_S, \hat{\mu}_{best} ) $
from all the realizations.
In this approach,
the $k$th 
bootstrap replication of
$n_{bg}$, $n^*_{bg}(k)$ is simulated 
by
sampling from a
Poisson distribution with expected value equal to 
$n_{bg}$. That is,
\begin{eqnarray}
n^*_{bg}(k)  \sim
Poi ( n_{bg}  ).
\end{eqnarray}
Given 
$n^*_{bg}(k)$,
the $k$th bootstrap replication of
$\hat{\mu}_B$,
$\hat{\mu}^*_B(k)$,
is
\begin{eqnarray}
\hat{\mu}^*_B(k) = \frac{ T } { T_{bg} } n^*_{bg}(k),
\end{eqnarray}
and
the $k$th bootstrap replication of
$\hat{\mu}_{best}$ is
$\hat{\mu}^*_{best}(k)= max(0,n-\hat{\mu}^{*}_B(k))$.
Thus, the $k$th bootstrap replication of
$P(n|\mu_S, \hat{\mu})$ is
\begin{eqnarray}
P(n|\mu_S, \hat{\mu}^*_B(k)) = (\mu_S + \hat{\mu}^*_B(k) ) ^n \frac{ \exp[ - ( \mu_S + \hat{\mu}^*_B(k) )  ] } { n !},
\end{eqnarray}
and
the $k$th bootstrap replication of
$P(n|\mu_S, \hat{\mu}_{best})$
is
\begin{eqnarray}
P(n|\mu_S, \hat{\mu}^*_{best}(k)) = (\mu_S + \hat{\mu}^*_{best}(k) ) ^n \frac{ \exp[ - ( \mu_S + \hat{\mu}^*_{best}(k) )  ] } { n !}.
\end{eqnarray}
From all $K$ bootstrap replications, we determine the following mean values
\begin{eqnarray}
\bar{P}(n|\mu_S, \hat{\mu}_B) =
\frac{1}{K}
\sum_{k=1}^{K}
P(n|\mu_S, \hat{\mu}^*_B(k))
\end{eqnarray}
and
\begin{eqnarray}
\bar{P}(n|\mu_S, \hat{\mu}_{best}) =
\frac{1}{K}
\sum_{k=1}^{K}
P(n|\mu_S, \hat{\mu}^*_{best}(k))
\end{eqnarray}
In this Monte Carlo implementation
of the RFC method, one replaces
${P}(n|\mu_S, \hat{\mu}_{B})$
and
${P}(n|\mu_S, \hat{\mu}_{best})$
with the right-hand sides of Eqns. 13 and 14.

To reduce computer run time,
we do not implement a Monte Carlo version of the RFC.
Instead,
we determine the left-hand sides of Eqns. 13 and 14 by
numerical integration.
For instance, we evaluate the left-hand side of Eq. 13 as
\begin{eqnarray}
\bar{P}(n|\mu_S, \hat{\mu}_B) =
\sum_{k=k_{low}}^{k_{hi}}
P(n|\mu_S, \hat{\mu}^*_B = \frac{ k T}{T_{bg} } ) w(k) 
\end{eqnarray}
where 
$w(k)$
is
\begin{eqnarray}
w(k) = 
\frac{  \exp( - n_{bg} ) n_{bg} ^ k  } {k!}.
\end{eqnarray}
To speed up the algorithm,
we select $k_{low}$ and $k_{hi}$ so that
the sum of the $w(k)$ terms
agrees with 1 to within approximately $10^{-8}$.
We use a similar method to determine the left-hand side of Eq. 14.

\subsection{Bayesian Method}
Following  [23],
we determine 
a Bayesian credibility interval for $\mu_S$
given measurements of $n_{obs}$ and $n_{bg}$.
In this approach, the priors $p(\mu_B)$
and
$p( \mu_S | \mu_{B})$
are both uniform from 0 to a large positive constant.
Results are presented for the limiting case where this positive constant
approaches infinity.
Based on a Bayes Theorem argument, one can show that
\begin{eqnarray}
p( \mu_S, \mu_B |  n_{obs}, n_{bg} ) \propto
p( \mu_B |  n_{bg} ) p( n_{obs} | \mu_S,\mu_B  ),
\end{eqnarray}
where
the posterior pdf for $\mu_B$
given $n_{bg}$ is
\begin{eqnarray}
p( \mu_B |  n_{bg}) =  
\TT 
\frac{
 \exp(- \TT \mu_{B})  
( \TT \mu_B)^{n_{bg} }
}
{ n_{bg} ! },
\end{eqnarray}
and
$p ( n_{obs} | \mu_S, \mu_B)$ is 
the Poisson likelihood function
\begin{eqnarray}
p ( n_{obs} | \mu_S, \mu_B) = \exp( -(\mu_S + \mu_B) ) ( \mu_S + \mu_B)^{n_{obs}} / n_{obs}!.
\end{eqnarray}
See Figure 2 for examples of Eq. 18.

Further,
marginalizing with respect to $\mu_B$, 
we get
\begin{eqnarray}
p( \mu_S | n_{obs}, n_{bg}) = 
\sum_{i=0}^{n_{obs}} C_i
\frac{  (\mu_S)^ i \exp( - \mu_S) } 
{ i!},
\end{eqnarray}
where
\begin{eqnarray}
C_i = \frac{
( 1 + \TT ) ^ i  ~ \frac{ (n_{obs}+ n_{bg} - i )!}{(n_{obs}-i)!}
}
{\sum_{j=0}^{n} 
( 1 + \TT ) ^ j  ~ \frac{ (n_{obs}+ n_{bg} - j )!}
{(n_{obs}-j)!}
}.
\end{eqnarray}
See [23] for more details of this derivation.

In this study, we determine the endpoints of 90 $\%$
and 95 $\%$
credibility intervals as follows.
For 
both the 90 $\%$
and 95 $\%$
cases, we determine the maximum lower endpoint of the one-sided interval $l_{max}$.
In an optimization code,
for each trial value of the lower endpoint $l$ (where
$ 0 \le l \le l_{max}$)
we determine 
the upper endpoint $u$ such that the integral of the posterior pdf
from $l$ to $u$ equals the nominal frequentist coverage.
We determine 
the lower endpoint $l$ that minimizes $u-l$.
If the optimal value of $l$ is less than 
the specified numerical tolerance ($10^{-6}$)
of the optimization algorithm,
we set it to 0.

\subsection{Propagation-of-Errors Method}
We compute two-sided confidence intervals with a
standard propagation-of-errors (POE) method that has a continuity correction.
The $1 - \alpha$ level  POE confidence 
interval for $\mu_S$ is $ \hat{\mu}_S \pm ( z_{\alpha/2} \hat{\sigma}_{\hat{\mu}_S} + 0.5) $ where
\begin{eqnarray}
\hat{\mu}_S =  n_{obs} - \hat{\mu}_B,
\end{eqnarray}
and
\begin{eqnarray}
\hat{\sigma}^2_{\hat{\mu}_S} = n_{obs} + (\frac{T}{T_{bg}})^2 n_{bg}.
\end{eqnarray}
For levels of 0.90 and 0.95, $z_{\alpha/2 } = $ 1.64 and 1.96, respectively.
We expect that this method will yield
confidence intervals with coverage close to
the desired nominal values
for the asymptotic case where the
signal-to-noise ratio of the data is high.
As a caveat, continuity corrections are typically
introduced when constructing confidence intervals for
the case where
there is no background [27]
rather than 
the more general case considered here.

In our simulation experiment, the POE method can yield
nonsensical results where one or both endpoints are negative
(Table 1). In our coverage studies,
we treat negative endpoints as 0. Hence, if
both endpoints are negative, the resultant interval
is treated as $(0,0)$.
In physics and astroparticle physics
experiments where
one hopes to discover a new particle,
null resuls are common and experimenters
provide upper limits.
If both endpoints are negative,
one can not set a reasonable upper limit.
Hence the POE method is clearly unacceptable
for low count
data sets.

\section{Simulation Experiments} 
In Table 1, we list some realizations of 
data and associated intervals constructed to have nominal coverage of 90 $\%$
for the four methods.  Based on 2000 realizations of data
for each of various choices of $\TT$ and $\mu_S$ and $\mu_B$,
we determine the frequentist coverage
as the 
fraction of the intervals that cover the true value of $\mu_S$ for each method
(Tables 2-7).
We also estimate detection probabilities for the different 
methods for levels 0.90 and 0.95 (Tables 8-13).

To start, we consider
90 percent intervals for the case where $\mu_B = 1$.
In Figures 3 and 4, we show coverage and detection probabilities
as a function of $\mu_S$ and $\TT$ for this case.
In general, when $\TT = 1$, the coverage properties of the FC and RFC 
methods are poor at low value of $\mu_S$.
In Figures 5 and 6, we show the false detection probabilies for
all cases for $\mu_S=0$. 
For $\TT = 1,2$, both the RFC and FC
methods have false detection probabilities that are
higher than predicted according to the nominal
frequentist coverage of the intervals.
Hence, reporting a discovery
based on an analysis with the FC or RFC method
should be treated with great caution for cases
where $\TT = 1,2$.
For $\TT \ge 5$, the false detection probabilities of the FC and
RFC methods are generally slightly less than their associated nominal
target values.
In contrast, for all values of $\TT$, the false detection probabilities
of the Bayesian method
are less than the values predicted by the nominal
frequentist coverage.
In Figures 7-10, we display coverage and
detection probabilities for all cases
considered in our simulation study.

In Tables 14 and 15, we list the root-mean-square (RMS) deviation between
the observed and nominal frequentist coverage
probabilities as a function of $\mu_B$.
We include results for $\mu_S \le 10$.
According to our coverage and detection probability criteria, the POE
method performed the least well of all methods.
This is not a surprise since
the poor performance of the POE method for low-count situations is well known.
In general, the Bayesian method yielded intervals
with the lowest detection probabilities compared to the 
FC and RFC methods.
According to the RMS coverage criterion,
the coverage properties of the Bayesian intervals
are inferior to the intervals produces by the FC and RFC methods
for most cases.
The exception to this pattern was for the case
where $\TT = 1$ and $\mu_B = 1,2$.
The 
FC and RFC method had better coverage compared to Bayesian method for  $\mu_B = 0.2$
for all values of  $\TT$ considered.

For $\TT=1,2$
the coverage properties of the RFC method where
slightly better than those of the FC method
for $\mu_B = 1,2$.
However, for $\TT$ greater than or equal to 5, the FC method
yields intervals with superior coverage and detection probabilities
compared to the RFC and Bayesian methods.

\subsection{Comments}

For fixed $\mu_B$, as $\mu_S $ increases, we sometimes
observe nonmonotic trends in coverage.
In other studies such as [24], 
nonmonotic trends were also observed.

We expect the FC method to yield
poor results
when the
1-sigma uncertainty in the
estimated background (Eq. 5)
is large.
For $\frac{ T_{bg}} { T }  = $ 1 and $\mu_B = $ 0.2, 1, 2,
Eq. 5 yields absolute uncertainties
of 0.45, 1 and 1.41,
and Eq. 6 yields fractional uncertainties of
224 $\%$, 100 $\%$ and 71 $\%$ respectively.
For $\frac{ T_{bg}} { T }  = $ 5,
the absolute and fractional uncertainties are
0.2, 0.45 and 0.63,
and
100 $\%$, 45 $\%$ and 31 $\%$ respectively.
For $\frac{ T_{bg}} { T }  = $ 25,
the absolute and fractional uncertainties
are 0.09, 0.20 and 0.28,
and 45 $\%$, 20 $\%$ and 14 $\%$ respectively.

It is plausible that the performance of
the FC method depends solely on the
Eq. 5
uncertainty of the background estimate.
However, comparison of the coverage properties
of the FC intervals 
for
the case where
$\frac{ T_{bg}} { T }  = $  1, $\mu_B = 0.2$
and
for the case where
$\frac{ T_{bg}} { T }  = $  5, $\mu_B = 1$,
suggest a more complicated picture.
For the first case, the
FC intervals have poor coverage at low values of $\mu_S$ (Tables 2,5).
For the second case,
the intervals have good coverage at all $\mu_S$ (Tables 3,6).
However, the standard deviation of the estimated background
is the same for both cases.
Perhaps this result is due to
differences in the shapes of the 
background estimate
pdfs for $\TT=1$ and $\TT = 5$.

In the POE method, we approximate the distribution of
the background-corrected estimate of $\mu_S$,
$n_{obs} - \hat{\mu}_B$, as
a normal (Gaussian) random variable.
For the special case where $\mu_B = $ 0,
a common rule of thumb is that the normal approximation is 
reasonable
when the expected value of $n_{obs}$ is 
greater than about
10 [27].
From this, we conclude that if the expected values of
of $n_{obs}$ and $n_{bg}$ both exceed 10,
the Gaussian assumption seems reasonable.
As a caveat, the adequacy of the normal approximation depends
on the goal of the analysis.
For instance, constructing a confidence interval with
nominal coverage of 0.99 is a more demanding task than constructing an interval
with nominal coverage of 0.90.
For the cases studied here where the nominal coverage 
is 0.90 or 0.95,
the POE intervals had coverage close to the desired nominal values when 
$\mu_S$ was greater than about 5.

In our implementation of the Bayesian method, we
specify uniform priors for $\mu_S$ and $\mu_B$
and
construct a  
minimum length credibility
interval.
Roe and Woodroofe [24]
determined a minimum length
credibility interval based on a uniform prior
for $\mu_S$
for the simpler problem where $\mu_B$ was assumed to be known.
Hence, our study can be regarded a generalization of [24] to
the case where $\mu_B$ is not known exactly.
As a caveat, in a Bayesian approach, one could
consider
other priors.
For
a given experiment, it is possible that other priors might be more appropriate
than the uniform
prior considered here.
How well such alternative Bayesian schemes would
perform relative to the one studied here is beyond the scope of this 
study.

As remarked earlier, we did not adjust our intervals to ensure that
the upper endpoint of the FC and RFC intervals are 
nondecreasing functions of $\mu_B$.
It is possible that 
such an adjustment
might improve the performance of the FC and RFC methods.
Also,
the FC method of computing
the likelihood-ratio term $R$ may not
be the best procedure
[28-31].

\section{Summary}

In this work,
we studied four methods to construct
uncertainty intervals for very weak Poisson signals
in the presence of background.
We considered the case where
a primary experiment
yielded a realization of the signal plus background,
and a second experiment yielded a realization of
the background.
The duration of the background-only experiment $T_{bg}$ and
and the duration of the primary experiment $T$ were selected so
that 
$\TT$ varied from 1 to 25.
This choice  of $\TT = $25
was
motivated by experimental studies
at NIST.
The values of the expected background $\mu_B$ varied from 0.2 to
2. The choice of the range was also motivated by NIST experiments.

We constructed confidence intervals
based on the standard propagation-of-errors method
as well as two implementations of a Neyman procedure
due to Feldman and Cousins (FC) and 
Conrad (RFC).
In the FC method, uncertainty in the background
was neglected. In our implementation of the RFC method, uncertainty
in the background
parameter was
accounted for.
In both of these methods, acceptance regions
were determined for each value of the expected signal
rate based on a likelihood-ratio
ordering principle. Hence,
the upper and lower endpoints of 
the confidence intervals were
automatically selected.
We also constructed minimum length Bayesian credibility intervals.

According to our coverage and detection probability criteria, the POE
method performed the least well of all methods.
In general, the Bayesian method yielded intervals
with the lowest detection probabilities compared to the 
FC and RFC methods (Tables 8-13, Figures 4,5,6,8 and 10).
According to an RMS criterion,
the coverage properties of the Bayesian intervals
were inferior to the intervals produces by the FC and RFC methods
(Tables 14 and 15) for most cases.
The exception to this pattern was for the case
where $\TT = 1$ and $\mu_B = 1,2$.

The 
FC and RFC methods had better coverage compared to the Bayesian method for  $\mu_B = 0.2$
for all values of  $\TT$ considered.
We expect similar results 
for $\mu_B < 0.2$.
We interpret this result as evidence that when expected number of
background counts is 0.2 or less, the FC method (which neglects 
uncertainty in the background) works well because uncertainty
in the observed background is not significant compared to
other sources of uncertainty that affect the interval.

For $\TT=1,2$
the coverage properties of the RFC method where
slightly better than those of the FC method
for $\mu_B = 1,2$.
However, for $\TT$ greater than or equal to 5, the FC method
yielded intervals with superior coverage and detection probabilities
compared to the RFC and Bayesian methods.
We attribute the  good performance of the FC
method to the fact that
uncertainty 
in the estimated background is not significant compared to
other sources of uncertainty that affect the interval
when $\mu_B \le 2$ and $\TT \ge 5 $.
The relative performance of the three methods for $\mu_B > 2$
is an open question. We speculate that for the FC method to yield a result superior to the
Bayesian or RFC method for $\mu_B$ much larger than 2, $\TT$ may have to
be larger than 5 in order to reduce the uncertainty of estimated background
to a sufficiently low level.

As a caveat, for $\TT = 1,2$, both the RFC and FC
methods had false detection probabilities that were
higher than predicted according to the nominal
frequentist coverage of the intervals for $\TT = 1,2$ (Figures 5,6).
Hence, reporting a discovery
based on an analysis with the FC or RFC method
should be treated with great caution for cases
where $\TT = 1,2$.
For $\TT \ge 5$, the false detection probabilities of the FC and
RFC methods were generally slightly less than their associated nominal
target values.
In contrast, for all values of $\TT$, the false detection probabilities
of the Bayesian method
were less than the value predicted by the nominal
frequentist coverage.

\begin{center}
{\bf Acknowledgements}
\end{center}
We thank H.K. Liu, L.A. Currie
and the anonymous reviewers of this work 
for useful comments.

\newpage{}
\begin {center}
\noindent {Table   1. Upper and lower endpoints of uncertainty intervals with
nominal frequentist coverage of 0.90.
The intervals are determined from simulated values of $n_{obs}$ and $n_{bg}$.
For informational purposes, we list $\mu_S$ and $\mu_B$. 
}
\\
\begin{tabular}{ccccccccccccc} \hline
\\
\\
& & & & &                     Bayesian&  Bayesian&  FC&      FC&     RFC&     RFC&    POE&    POE
\\
$\frac{T_{bg}}{T}$ &  $\mu_ S$ &  $\mu_B$ &     $n_{obs}$ &  $n_{bg}$ & Lower &    Upper &    Lower &   Upper&  Lower&   Upper&  Lower&  Upper
\\
\\
   1&    1.0&    2.0&  2&  1&     0.00&     4.32 &     0.00&     4.91 &    0.00&     5.27 &   -2.35&     4.35
\\
   1&   10.0&    0.2&  6&  0&     1.58&    10.42 &    2.21&    11.46  &   2.21&    11.46  &   1.47&    10.53
\\
\\
   5&    2.0&    0.2&  1&  0&     0.00&     3.74 &    0.11&     4.35  &   0.11&     4.35  &  -1.14&     3.14
\\
   5&    1.0&    1.0&  1&  4&     0.00&     3.34 &    0.00&     3.55  &   0.00&     3.56  &  -2.07&     2.47
\\
   5&    5.0&    2.0&  5&  7&     0.49&     8.11 &    1.04&     8.58  &   0.95&     8.58  &  -0.68&     7.88
\\
\\
  25&    0.1&    2.0&  0& 46&     0.00&     2.30  &   0.00&     1.15  &   0.00&     1.15  &  -2.79&    -0.89
\\
  25&    5.0&    1.0&  2& 16&     0.00&     4.70  &   0.00&     5.27  &   0.00&     5.27  &  -1.48&     4.20
\\
  25&   10.0&    0.2&  9&  7&     4.57&    14.62  &   4.08&    15.01  &   4.08&    15.04  &   3.28&    14.16
\\ \hline
\end {tabular}
\end{center}
\newpage{}

\begin {center}
\noindent {Table   2.
Estimated coverage probabilities of uncertainty intervals with
nominal frequentist coverage of 0.90.
Approximate 68 percent uncertainty
due to sampling variability given for cases where estimated coverage probability
is greater than 0 or less than 1.
$\frac{T_{bg}}{T} = 1$.
}
\\
\begin{tabular}{cccccc} \hline
\\
\\
 $\mu_ S$ &  $\mu_B$ & Bayesian     & FC &    RFC &   POE
\\
\\
    0.0&    0.2&       1&       0.857$\pm$       0.008&       0.857$\pm$       0.008&       1
\\
    0.1&    0.2&       1&       0.797$\pm$       0.009&       0.798$\pm$       0.009&       1
\\
    0.2&    0.2&       1&       0.948$\pm$       0.005&       0.948$\pm$       0.005&       1
\\
    1.0&    0.2&       0.998$\pm$       0.001&       0.908$\pm$       0.006&       0.908$\pm$       0.006&       0.751$\pm$       0.010
\\
    2.0&    0.2&       0.995$\pm$       0.002&       0.960$\pm$       0.004&       0.961$\pm$       0.004&       0.887$\pm$       0.007
\\
    5.0&    0.2&       0.867$\pm$       0.008&       0.923$\pm$       0.006&       0.933$\pm$       0.006&       0.869$\pm$       0.008
\\
   10.0&    0.2&       0.906$\pm$       0.007&       0.904$\pm$       0.007&       0.906$\pm$       0.007&       0.907$\pm$       0.007
\\
   20.0&    0.2&       0.911$\pm$       0.006&       0.904$\pm$       0.007&       0.908$\pm$       0.006&       0.911$\pm$       0.006
\\
\\
    0.0&    1.0&       0.990$\pm$       0.002&       0.728$\pm$       0.010&       0.748$\pm$       0.010&       0.991$\pm$       0.002
\\
    0.1&    1.0&       0.992$\pm$       0.002&       0.708$\pm$       0.010&       0.733$\pm$       0.010&       0.985$\pm$       0.003
\\
    0.2&    1.0&       0.986$\pm$       0.003&       0.861$\pm$       0.008&       0.862$\pm$       0.008&       0.982$\pm$       0.003
\\
    1.0&    1.0&       0.994$\pm$       0.002&       0.841$\pm$       0.008&       0.856$\pm$       0.008&       0.906$\pm$       0.007
\\
    2.0&    1.0&       0.981$\pm$       0.003&       0.909$\pm$       0.006&       0.927$\pm$       0.006&       0.924$\pm$       0.006
\\
    5.0&    1.0&       0.900$\pm$       0.007&       0.901$\pm$       0.007&       0.922$\pm$       0.006&       0.916$\pm$       0.006
\\
   10.0&    1.0&       0.896$\pm$       0.007&       0.906$\pm$       0.007&       0.921$\pm$       0.006&       0.917$\pm$       0.006
\\
   20.0&    1.0&       0.903$\pm$       0.007&       0.886$\pm$       0.007&       0.901$\pm$       0.007&       0.903$\pm$       0.007
\\
\\
    0.0&    2.0&       0.964$\pm$       0.004&       0.786$\pm$       0.009&       0.835$\pm$       0.008&       0.974$\pm$       0.004
\\
    0.1&    2.0&       0.972$\pm$       0.004&       0.763$\pm$       0.010&       0.828$\pm$       0.008&       0.949$\pm$       0.005
\\
    0.2&    2.0&       0.968$\pm$       0.004&       0.843$\pm$       0.008&       0.857$\pm$       0.008&       0.951$\pm$       0.005
\\
    1.0&    2.0&       0.980$\pm$       0.003&       0.838$\pm$       0.008&       0.867$\pm$       0.008&       0.939$\pm$       0.005
\\
    2.0&    2.0&       0.978$\pm$       0.003&       0.866$\pm$       0.008&       0.930$\pm$       0.006&       0.926$\pm$       0.006
\\
    5.0&    2.0&       0.910$\pm$       0.006&       0.886$\pm$       0.007&       0.927$\pm$       0.006&       0.924$\pm$       0.006
\\
   10.0&    2.0&       0.872$\pm$       0.007&       0.872$\pm$       0.007&       0.912$\pm$       0.006&       0.913$\pm$       0.006
\\
   20.0&    2.0&       0.907$\pm$       0.007&       0.886$\pm$       0.007&       0.908$\pm$       0.006&       0.910$\pm$       0.006
\\ \hline
\end {tabular}
\end{center}
\newpage{}
\newpage{}
\newpage{}
\newpage{}

\begin {center}
\noindent {Table   3. Estimated coverage probabilities of uncertainty intervals with
nominal frequentist coverage of 0.90.
Approximate 68 percent uncertainty
due to sampling variability given for cases where estimated coverage probability
is greater than 0 or less than 1.
$\frac{T_{bg}}{T} = 5$.
}
\\
\begin{tabular}{cccccc} \hline
\\
\\
 $\mu_ S$ &  $\mu_B$ & Bayesian     & FC &    RFC &   POE
\\
\\
    0.0&    0.2&       0.993$\pm$       0.002&       0.928$\pm$       0.006&       0.928$\pm$       0.006&       1
\\
    0.1&    0.2&       0.981$\pm$       0.003&       0.882$\pm$       0.007&       0.888$\pm$       0.007&       1
\\
    0.2&    0.2&       0.974$\pm$       0.004&       0.955$\pm$       0.005&       0.955$\pm$       0.005&       0.999$\pm$       0.001
\\
    1.0&    0.2&       0.971$\pm$       0.004&       0.934$\pm$       0.006&       0.934$\pm$       0.006&       0.682$\pm$       0.010
\\
    2.0&    0.2&       0.973$\pm$       0.004&       0.967$\pm$       0.004&       0.967$\pm$       0.004&       0.866$\pm$       0.008
\\
    5.0&    0.2&       0.909$\pm$       0.006&       0.930$\pm$       0.006&       0.930$\pm$       0.006&       0.868$\pm$       0.008
\\
   10.0&    0.2&       0.892$\pm$       0.007&       0.901$\pm$       0.007&       0.901$\pm$       0.007&       0.912$\pm$       0.006
\\
   20.0&    0.2&       0.892$\pm$       0.007&       0.894$\pm$       0.007&       0.894$\pm$       0.007&       0.907$\pm$       0.006
\\
\\
    0.0&    1.0&       0.964$\pm$       0.004&       0.924$\pm$       0.006&       0.924$\pm$       0.006&       0.997$\pm$       0.001
\\
    0.1&    1.0&       0.956$\pm$       0.005&       0.911$\pm$       0.006&       0.939$\pm$       0.005&       0.916$\pm$       0.006
\\
    0.2&    1.0&       0.969$\pm$       0.004&       0.929$\pm$       0.006&       0.929$\pm$       0.006&       0.872$\pm$       0.007
\\
    1.0&    1.0&       0.967$\pm$       0.004&       0.936$\pm$       0.005&       0.937$\pm$       0.005&       0.850$\pm$       0.008
\\
    2.0&    1.0&       0.968$\pm$       0.004&       0.912$\pm$       0.006&       0.915$\pm$       0.006&       0.902$\pm$       0.007
\\
    5.0&    1.0&       0.912$\pm$       0.006&       0.929$\pm$       0.006&       0.929$\pm$       0.006&       0.922$\pm$       0.006
\\
   10.0&    1.0&       0.893$\pm$       0.007&       0.909$\pm$       0.006&       0.909$\pm$       0.006&       0.909$\pm$       0.006
\\
   20.0&    1.0&       0.891$\pm$       0.007&       0.897$\pm$       0.007&       0.902$\pm$       0.007&       0.901$\pm$       0.007
\\
\\
    0.0&    2.0&       0.944$\pm$       0.005&       0.906$\pm$       0.007&       0.908$\pm$       0.006&       0.996$\pm$       0.001
\\
    0.1&    2.0&       0.954$\pm$       0.005&       0.912$\pm$       0.006&       0.938$\pm$       0.005&       0.877$\pm$       0.007
\\
    0.2&    2.0&       0.944$\pm$       0.005&       0.902$\pm$       0.007&       0.907$\pm$       0.006&       0.884$\pm$       0.007
\\
    1.0&    2.0&       0.963$\pm$       0.004&       0.910$\pm$       0.006&       0.922$\pm$       0.006&       0.920$\pm$       0.006
\\
    2.0&    2.0&       0.966$\pm$       0.004&       0.921$\pm$       0.006&       0.926$\pm$       0.006&       0.902$\pm$       0.007
\\
    5.0&    2.0&       0.901$\pm$       0.007&       0.919$\pm$       0.006&       0.920$\pm$       0.006&       0.910$\pm$       0.006
\\
   10.0&    2.0&       0.902$\pm$       0.007&       0.909$\pm$       0.006&       0.913$\pm$       0.006&       0.914$\pm$       0.006
\\
   20.0&    2.0&       0.897$\pm$       0.007&       0.902$\pm$       0.007&       0.905$\pm$       0.007&       0.908$\pm$       0.006
\\ \hline
\end {tabular}
\end{center}

\newpage{}
\begin {center}
\noindent {Table   4. Estimated coverage probabilities of uncertainty intervals with
nominal frequentist coverage of 0.90.
Approximate 68 percent uncertainty
due to sampling variability given for cases where estimated coverage probability
is greater than 0 or less than 1.
$\frac{T_{bg}}{T} = 25$.
}
\\
\begin{tabular}{cccccc} \hline
\\
\\
 $\mu_ S$ &  $\mu_B$ & Bayesian     & FC &    RFC &   POE
\\
\\
    0.0&    0.2&       0.982$\pm$       0.003&       0.973$\pm$       0.004&       0.973$\pm$       0.004&       1
\\
    0.1&    0.2&       0.974$\pm$       0.004&       0.967$\pm$       0.004&       0.967$\pm$       0.004&       1
\\
    0.2&    0.2&       0.953$\pm$       0.005&       0.937$\pm$       0.005&       0.937$\pm$       0.005&       1
\\
    1.0&    0.2&       0.969$\pm$       0.004&       0.954$\pm$       0.005&       0.954$\pm$       0.005&       0.698$\pm$       0.010
\\
    2.0&    0.2&       0.974$\pm$       0.004&       0.979$\pm$       0.003&       0.979$\pm$       0.003&       0.884$\pm$       0.007
\\
    5.0&    0.2&       0.925$\pm$       0.006&       0.928$\pm$       0.006&       0.928$\pm$       0.006&       0.891$\pm$       0.007
\\
   10.0&    0.2&       0.890$\pm$       0.007&       0.906$\pm$       0.007&       0.906$\pm$       0.007&       0.920$\pm$       0.006
\\
   20.0&    0.2&       0.906$\pm$       0.007&       0.909$\pm$       0.006&       0.909$\pm$       0.006&       0.920$\pm$       0.006
\\
\\
    0.0&    1.0&       0.957$\pm$       0.005&       0.940$\pm$       0.005&       0.940$\pm$       0.005&       0.999$\pm$       0.001
\\
    0.1&    1.0&       0.958$\pm$       0.005&       0.931$\pm$       0.006&       0.939$\pm$       0.005&       0.665$\pm$       0.011
\\
    0.2&    1.0&       0.960$\pm$       0.004&       0.931$\pm$       0.006&       0.931$\pm$       0.006&       0.709$\pm$       0.010
\\
    1.0&    1.0&       0.961$\pm$       0.004&       0.946$\pm$       0.005&       0.946$\pm$       0.005&       0.875$\pm$       0.007
\\
    2.0&    1.0&       0.961$\pm$       0.004&       0.917$\pm$       0.006&       0.917$\pm$       0.006&       0.916$\pm$       0.006
\\
    5.0&    1.0&       0.900$\pm$       0.007&       0.927$\pm$       0.006&       0.927$\pm$       0.006&       0.927$\pm$       0.006
\\
   10.0&    1.0&       0.910$\pm$       0.006&       0.930$\pm$       0.006&       0.930$\pm$       0.006&       0.914$\pm$       0.006
\\
   20.0&    1.0&       0.907$\pm$       0.006&       0.914$\pm$       0.006&       0.914$\pm$       0.006&       0.918$\pm$       0.006
\\
\\
    0.0&    2.0&       0.948$\pm$       0.005&       0.929$\pm$       0.006&       0.934$\pm$       0.006&       0.998$\pm$       0.001
\\
    0.1&    2.0&       0.946$\pm$       0.005&       0.930$\pm$       0.006&       0.933$\pm$       0.006&       0.873$\pm$       0.007
\\
    0.2&    2.0&       0.949$\pm$       0.005&       0.928$\pm$       0.006&       0.928$\pm$       0.006&       0.889$\pm$       0.007
\\
    1.0&    2.0&       0.957$\pm$       0.005&       0.919$\pm$       0.006&       0.924$\pm$       0.006&       0.909$\pm$       0.006
\\
    2.0&    2.0&       0.957$\pm$       0.005&       0.930$\pm$       0.006&       0.931$\pm$       0.006&       0.902$\pm$       0.007
\\
    5.0&    2.0&       0.881$\pm$       0.007&       0.913$\pm$       0.006&       0.913$\pm$       0.006&       0.896$\pm$       0.007
\\
   10.0&    2.0&       0.892$\pm$       0.007&       0.908$\pm$       0.006&       0.909$\pm$       0.006&       0.897$\pm$       0.007
\\
   20.0&    2.0&       0.900$\pm$       0.007&       0.909$\pm$       0.006&       0.909$\pm$       0.006&       0.912$\pm$       0.006
\\ \hline
\end {tabular}
\end{center}
\newpage{}
\begin {center}
\noindent {Table   5. Estimated coverage probabilities of uncertainty intervals with
nominal frequentist coverage of 0.95.
Approximate 68 percent uncertainty
due to sampling variability given for cases where estimated coverage probability
is greater than 0 or less than 1.
$\frac{T_{bg}}{T} = 1$.
}
\\
\begin{tabular}{cccccc} \hline
\\
\\
 $\mu_ S$ &  $\mu_B$ & Bayesian     & FC &    RFC &   POE
\\
\\
    0.0&    0.2&       1&       0.857$\pm$       0.008&       0.857$\pm$       0.008&       1
\\
    0.1&    0.2&       1&       0.976$\pm$       0.003&       0.976$\pm$       0.003&       1
\\
    0.2&    0.2&       1&       0.948$\pm$       0.005&       0.948$\pm$       0.005&       1
\\
    1.0&    0.2&       1&       0.974$\pm$       0.004&       0.974$\pm$       0.004&       0.755$\pm$       0.010
\\
    2.0&    0.2&       0.998$\pm$       0.001&       0.976$\pm$       0.003&       0.977$\pm$       0.003&       0.892$\pm$       0.007
\\
    5.0&    0.2&       0.962$\pm$       0.004&       0.972$\pm$       0.004&       0.974$\pm$       0.004&       0.953$\pm$       0.005
\\
   10.0&    0.2&       0.951$\pm$       0.005&       0.949$\pm$       0.005&       0.949$\pm$       0.005&       0.935$\pm$       0.006
\\
   20.0&    0.2&       0.949$\pm$       0.005&       0.953$\pm$       0.005&       0.955$\pm$       0.005&       0.950$\pm$       0.005
\\
\\
    0.0&    1.0&       0.999$\pm$       0.001&       0.748$\pm$       0.010&       0.755$\pm$       0.010&       0.999$\pm$       0.001
\\
    0.1&    1.0&       0.998$\pm$       0.001&       0.878$\pm$       0.007&       0.886$\pm$       0.007&       0.995$\pm$       0.002
\\
    0.2&    1.0&       0.997$\pm$       0.001&       0.862$\pm$       0.008&       0.870$\pm$       0.008&       0.998$\pm$       0.001
\\
    1.0&    1.0&       0.998$\pm$       0.001&       0.945$\pm$       0.005&       0.945$\pm$       0.005&       0.940$\pm$       0.005
\\
    2.0&    1.0&       0.993$\pm$       0.002&       0.949$\pm$       0.005&       0.956$\pm$       0.005&       0.946$\pm$       0.005
\\
    5.0&    1.0&       0.967$\pm$       0.004&       0.951$\pm$       0.005&       0.969$\pm$       0.004&       0.958$\pm$       0.004
\\
   10.0&    1.0&       0.958$\pm$       0.004&       0.959$\pm$       0.004&       0.962$\pm$       0.004&       0.960$\pm$       0.004
\\
   20.0&    1.0&       0.954$\pm$       0.005&       0.944$\pm$       0.005&       0.955$\pm$       0.005&       0.950$\pm$       0.005
\\
\\
    0.0&    2.0&       0.993$\pm$       0.002&       0.835$\pm$       0.008&       0.864$\pm$       0.008&       0.994$\pm$       0.002
\\
    0.1&    2.0&       0.988$\pm$       0.002&       0.869$\pm$       0.008&       0.900$\pm$       0.007&       0.981$\pm$       0.003
\\
    0.2&    2.0&       0.990$\pm$       0.002&       0.856$\pm$       0.008&       0.887$\pm$       0.007&       0.989$\pm$       0.002
\\
    1.0&    2.0&       0.993$\pm$       0.002&       0.921$\pm$       0.006&       0.928$\pm$       0.006&       0.965$\pm$       0.004
\\
    2.0&    2.0&       0.991$\pm$       0.002&       0.935$\pm$       0.006&       0.948$\pm$       0.005&       0.966$\pm$       0.004
\\
    5.0&    2.0&       0.967$\pm$       0.004&       0.937$\pm$       0.005&       0.957$\pm$       0.005&       0.962$\pm$       0.004
\\
   10.0&    2.0&       0.944$\pm$       0.005&       0.943$\pm$       0.005&       0.958$\pm$       0.004&       0.958$\pm$       0.004
\\
   20.0&    2.0&       0.952$\pm$       0.005&       0.945$\pm$       0.005&       0.957$\pm$       0.005&       0.958$\pm$       0.005
\\ \hline
\end {tabular}
\end{center}
\newpage{}

\begin {center}
\noindent {Table   6. Estimated coverage probabilities of uncertainty intervals with
nominal frequentist coverage of 0.95.
Approximate 68 percent uncertainty
due to sampling variability given for cases where estimated coverage probability
is greater than 0 or less than 1.
$\frac{T_{bg}}{T} = 5$.
}
\\
\begin{tabular}{cccccc} \hline
\\
\\
 $\mu_ S$ &  $\mu_B$ & Bayesian     & FC &    RFC &   POE
\\
    0.0&    0.2&       0.997$\pm$       0.001&       0.932$\pm$       0.006&       0.932$\pm$       0.006&       1
\\
    0.1&    0.2&       0.997$\pm$       0.001&       0.972$\pm$       0.004&       0.972$\pm$       0.004&       1
\\
    0.2&    0.2&       0.993$\pm$       0.002&       0.974$\pm$       0.004&       0.974$\pm$       0.004&       1
\\
    1.0&    0.2&       0.990$\pm$       0.002&       0.971$\pm$       0.004&       0.971$\pm$       0.004&       0.684$\pm$       0.010
\\
    2.0&    0.2&       0.983$\pm$       0.003&       0.976$\pm$       0.003&       0.976$\pm$       0.003&       0.871$\pm$       0.007
\\
    5.0&    0.2&       0.950$\pm$       0.005&       0.962$\pm$       0.004&       0.962$\pm$       0.004&       0.936$\pm$       0.005
\\
   10.0&    0.2&       0.949$\pm$       0.005&       0.950$\pm$       0.005&       0.950$\pm$       0.005&       0.933$\pm$       0.006
\\
   20.0&    0.2&       0.945$\pm$       0.005&       0.951$\pm$       0.005&       0.955$\pm$       0.005&       0.950$\pm$       0.005
\\
\\
    0.0&    1.0&       0.983$\pm$       0.003&       0.952$\pm$       0.005&       0.962$\pm$       0.004&       1
\\
    0.1&    1.0&       0.983$\pm$       0.003&       0.953$\pm$       0.005&       0.955$\pm$       0.005&       0.955$\pm$       0.005
\\
    0.2&    1.0&       0.986$\pm$       0.003&       0.953$\pm$       0.005&       0.969$\pm$       0.004&       0.921$\pm$       0.006
\\
    1.0&    1.0&       0.984$\pm$       0.003&       0.967$\pm$       0.004&       0.967$\pm$       0.004&       0.855$\pm$       0.008
\\
    2.0&    1.0&       0.982$\pm$       0.003&       0.968$\pm$       0.004&       0.969$\pm$       0.004&       0.930$\pm$       0.006
\\
    5.0&    1.0&       0.961$\pm$       0.004&       0.966$\pm$       0.004&       0.966$\pm$       0.004&       0.945$\pm$       0.005
\\
   10.0&    1.0&       0.950$\pm$       0.005&       0.959$\pm$       0.004&       0.959$\pm$       0.004&       0.955$\pm$       0.005
\\
   20.0&    1.0&       0.946$\pm$       0.005&       0.946$\pm$       0.005&       0.950$\pm$       0.005&       0.941$\pm$       0.005
\\
\\
    0.0&    2.0&       0.980$\pm$       0.003&       0.939$\pm$       0.005&       0.957$\pm$       0.005&       0.999$\pm$       0.001
\\
    0.1&    2.0&       0.980$\pm$       0.003&       0.954$\pm$       0.005&       0.969$\pm$       0.004&       0.895$\pm$       0.007
\\
    0.2&    2.0&       0.978$\pm$       0.003&       0.943$\pm$       0.005&       0.948$\pm$       0.005&       0.902$\pm$       0.007
\\
    1.0&    2.0&       0.977$\pm$       0.003&       0.951$\pm$       0.005&       0.963$\pm$       0.004&       0.945$\pm$       0.005
\\
    2.0&    2.0&       0.985$\pm$       0.003&       0.959$\pm$       0.004&       0.965$\pm$       0.004&       0.927$\pm$       0.006
\\
    5.0&    2.0&       0.960$\pm$       0.004&       0.957$\pm$       0.005&       0.958$\pm$       0.005&       0.943$\pm$       0.005
\\
   10.0&    2.0&       0.945$\pm$       0.005&       0.959$\pm$       0.004&       0.959$\pm$       0.004&       0.954$\pm$       0.005
\\
   20.0&    2.0&       0.944$\pm$       0.005&       0.947$\pm$       0.005&       0.949$\pm$       0.005&       0.947$\pm$       0.005
\\ \hline
\end {tabular}
\end{center}

\newpage{}

\begin {center}
\noindent {Table   7. Estimated coverage probabilities of uncertainty intervals with
nominal frequentist coverage of 0.95.
Approximate 68 percent uncertainty
due to sampling variability given for cases where estimated coverage probability
is greater than 0 or less than 1.
$\frac{T_{bg}}{T} = 25$.
}
\\
\begin{tabular}{cccccc} \hline
\\
\\
 $\mu_ S$ &  $\mu_B$ & Bayesian     & FC &    RFC &   POE
\\
\\
    0.0&    0.2&       0.990$\pm$       0.002&       0.983$\pm$       0.003&       0.983$\pm$       0.003&       1
\\
    0.1&    0.2&       0.987$\pm$       0.003&       0.978$\pm$       0.003&       0.978$\pm$       0.003&       1
\\
    0.2&    0.2&       0.990$\pm$       0.002&       0.979$\pm$       0.003&       0.979$\pm$       0.003&       1
\\
    1.0&    0.2&       0.984$\pm$       0.003&       0.970$\pm$       0.004&       0.970$\pm$       0.004&       0.700$\pm$       0.010
\\
    2.0&    0.2&       0.982$\pm$       0.003&       0.981$\pm$       0.003&       0.981$\pm$       0.003&       0.886$\pm$       0.007
\\
    5.0&    0.2&       0.949$\pm$       0.005&       0.964$\pm$       0.004&       0.964$\pm$       0.004&       0.948$\pm$       0.005
\\
   10.0&    0.2&       0.956$\pm$       0.005&       0.956$\pm$       0.005&       0.956$\pm$       0.005&       0.934$\pm$       0.006
\\
   20.0&    0.2&       0.948$\pm$       0.005&       0.956$\pm$       0.005&       0.956$\pm$       0.005&       0.958$\pm$       0.005
\\
\\
    0.0&    1.0&       0.981$\pm$       0.003&       0.971$\pm$       0.004&       0.972$\pm$       0.004&       1
\\
    0.1&    1.0&       0.978$\pm$       0.003&       0.970$\pm$       0.004&       0.970$\pm$       0.004&       0.687$\pm$       0.010
\\
    0.2&    1.0&       0.983$\pm$       0.003&       0.969$\pm$       0.004&       0.973$\pm$       0.004&       0.715$\pm$       0.010
\\
    1.0&    1.0&       0.984$\pm$       0.003&       0.967$\pm$       0.004&       0.970$\pm$       0.004&       0.876$\pm$       0.007
\\
    2.0&    1.0&       0.981$\pm$       0.003&       0.980$\pm$       0.003&       0.980$\pm$       0.003&       0.949$\pm$       0.005
\\
    5.0&    1.0&       0.966$\pm$       0.004&       0.966$\pm$       0.004&       0.966$\pm$       0.004&       0.937$\pm$       0.005
\\
   10.0&    1.0&       0.955$\pm$       0.005&       0.972$\pm$       0.004&       0.972$\pm$       0.004&       0.965$\pm$       0.004
\\
   20.0&    1.0&       0.956$\pm$       0.005&       0.956$\pm$       0.005&       0.956$\pm$       0.005&       0.954$\pm$       0.005
\\
\\
    0.0&    2.0&       0.978$\pm$       0.003&       0.964$\pm$       0.004&       0.969$\pm$       0.004&        1
\\
    0.1&    2.0&       0.973$\pm$       0.004&       0.964$\pm$       0.004&       0.967$\pm$       0.004&       0.877$\pm$       0.007
\\
    0.2&    2.0&       0.979$\pm$       0.003&       0.964$\pm$       0.004&       0.966$\pm$       0.004&       0.894$\pm$       0.007
\\
    1.0&    2.0&       0.981$\pm$       0.003&       0.964$\pm$       0.004&       0.967$\pm$       0.004&       0.946$\pm$       0.005
\\
    2.0&    2.0&       0.981$\pm$       0.003&       0.960$\pm$       0.004&       0.961$\pm$       0.004&       0.911$\pm$       0.006
\\
    5.0&    2.0&       0.940$\pm$       0.005&       0.950$\pm$       0.005&       0.949$\pm$       0.005&       0.929$\pm$       0.006
\\
   10.0&    2.0&       0.942$\pm$       0.005&       0.957$\pm$       0.005&       0.957$\pm$       0.005&       0.941$\pm$       0.005
\\
   20.0&    2.0&       0.946$\pm$       0.005&       0.952$\pm$       0.005&       0.952$\pm$       0.005&       0.948$\pm$       0.005
\\ \hline
\end {tabular}
\end{center}
\newpage{}

\begin {center}
\noindent {Table   8. Estimated detection probabilities corresponding to uncertainty intervals with
nominal frequentist coverage of 0.90.
Approximate 68 percent uncertainty
due to sampling variability given for cases where estimated coverage probability
is greater than 0 or less than 1.
$\frac{T_{bg}}{T} = 1$.
}
\\
\begin{tabular}{cccccc} \hline
\\
\\
 $\mu_ S$ &  $\mu_B$ & Bayesian     & FC &    RFC &   POE
\\
\\
    0.0&    0.2&       0&       0.143$\pm$       0.008&       0.143$\pm$       0.008&       0
\\
    0.1&    0.2&       0&       0.203$\pm$       0.009&       0.202$\pm$       0.009&       0
\\
    0.2&    0.2&       0.001$\pm$     0.001&       0.269$\pm$       0.010&       0.268$\pm$       0.010&       0.001$\pm$       0.001
\\
    1.0&    0.2&       0.028$\pm$       0.004&       0.591$\pm$       0.011&       0.580$\pm$       0.011&       0.027$\pm$       0.004
\\
    2.0&    0.2&       0.152$\pm$       0.008&       0.795$\pm$       0.009&       0.759$\pm$       0.010&       0.144$\pm$       0.008
\\
    5.0&    0.2&       0.725$\pm$       0.010&       0.975$\pm$       0.004&       0.951$\pm$       0.005&       0.695$\pm$       0.010
\\
   10.0&    0.2&       0.979$\pm$       0.003&       1  &       0.998$\pm$       0.001&       0.971$\pm$       0.004
\\
   20.0&    0.2&       1&       1&       1&       1
\\
\\
    0.0&    1.0&       0.010$\pm$       0.002&       0.272$\pm$       0.010&       0.252$\pm$       0.010&       0.009$\pm$       0.002
\\
    0.1&    1.0&       0.010$\pm$       0.002&       0.293$\pm$       0.010&       0.268$\pm$       0.010&       0.009$\pm$       0.002
\\
    0.2&    1.0&       0.016$\pm$       0.003&       0.310$\pm$       0.010&       0.280$\pm$       0.010&       0.015$\pm$       0.003
\\
    1.0&    1.0&       0.070$\pm$       0.006&       0.461$\pm$       0.011&       0.381$\pm$       0.011&       0.055$\pm$       0.005
\\
    2.0&    1.0&       0.193$\pm$       0.009&       0.602$\pm$       0.011&       0.495$\pm$       0.011&       0.157$\pm$       0.008
\\
    5.0&    1.0&       0.673$\pm$       0.010&       0.887$\pm$       0.007&       0.819$\pm$       0.009&       0.571$\pm$       0.011
\\
   10.0&    1.0&       0.967$\pm$       0.004&       0.994$\pm$       0.002&       0.985$\pm$       0.003&       0.936$\pm$       0.005
\\
   20.0&    1.0&       1&       1&       1&       1
\\
\\
    0.0&    2.0&       0.036$\pm$       0.004&       0.214$\pm$       0.009&       0.165$\pm$       0.008&       0.026$\pm$       0.004
\\
    0.1&    2.0&       0.039$\pm$       0.004&       0.238$\pm$       0.010&       0.172$\pm$       0.008&       0.028$\pm$       0.004
\\
    0.2&    2.0&       0.043$\pm$       0.005&       0.253$\pm$       0.010&       0.183$\pm$       0.009&       0.031$\pm$       0.004
\\
    1.0&    2.0&       0.106$\pm$       0.007&       0.343$\pm$       0.011&       0.249$\pm$       0.010&       0.073$\pm$       0.006
\\
    2.0&    2.0&       0.228$\pm$       0.009&       0.493$\pm$       0.011&       0.374$\pm$       0.011&       0.162$\pm$       0.008
\\
    5.0&    2.0&       0.586$\pm$       0.011&       0.805$\pm$       0.009&       0.702$\pm$       0.010&       0.461$\pm$       0.011
\\
   10.0&    2.0&       0.918$\pm$       0.006&       0.977$\pm$       0.003&       0.949$\pm$       0.005&       0.850$\pm$       0.008
\\
   20.0&    2.0&       0.999$\pm$       0.001&       1&       0.999$\pm$       0.001&       0.997$\pm$       0.001
\\ \hline
\end {tabular}
\end{center}
\newpage{}
\begin {center}
\noindent {Table   9. Estimated detection probabilities corresponding to uncertainty intervals with
nominal frequentist coverage of 0.90.
Approximate 68 percent uncertainty
due to sampling variability given for cases where estimated coverage probability
is greater than 0 or less than 1.
$\frac{T_{bg}}{T} = 5$.
}
\\
\begin{tabular}{cccccc} \hline
\\
\\
 $\mu_ S$ &  $\mu_B$ & Bayesian     & FC &    RFC &   POE
\\
\\
    0.0&    0.2&       0.007$\pm$       0.002&       0.072$\pm$       0.006&       0.072$\pm$       0.006&       0
\\
    0.1&    0.2&       0.029$\pm$       0.004&       0.118$\pm$       0.007&       0.118$\pm$       0.007&       0
\\
    0.2&    0.2&       0.045$\pm$       0.005&       0.148$\pm$       0.008&       0.148$\pm$       0.008&       0.001$\pm$       0.001
\\
    1.0&    0.2&       0.281$\pm$       0.010&       0.453$\pm$       0.011&       0.453$\pm$       0.011&       0.022$\pm$       0.003
\\
    2.0&    0.2&       0.549$\pm$       0.011&       0.694$\pm$       0.010&       0.694$\pm$       0.010&       0.119$\pm$       0.007
\\
    5.0&    0.2&       0.935$\pm$       0.006&       0.962$\pm$       0.004&       0.962$\pm$       0.004&       0.654$\pm$       0.011
\\
   10.0&    0.2&       0.998$\pm$       0.001&       0.999$\pm$       0.001&       0.999$\pm$       0.001&       0.977$\pm$       0.003
\\
   20.0&    0.2&       1&       1&       1&       1
\\
\\
    0.0&    1.0&       0.036$\pm$       0.004&       0.076$\pm$       0.006&       0.076$\pm$       0.006&       0.003$\pm$       0.001
\\
    0.1&    1.0&       0.048$\pm$       0.005&       0.089$\pm$       0.006&       0.089$\pm$       0.006&       0.005$\pm$       0.001
\\
    0.2&    1.0&       0.055$\pm$       0.005&       0.103$\pm$       0.007&       0.103$\pm$       0.007&       0.003$\pm$       0.001
\\
    1.0&    1.0&       0.176$\pm$       0.009&       0.284$\pm$       0.010&       0.283$\pm$       0.010&       0.023$\pm$       0.003
\\
    2.0&    1.0&       0.381$\pm$       0.011&       0.509$\pm$       0.011&       0.509$\pm$       0.011&       0.114$\pm$       0.007
\\
    5.0&    1.0&       0.852$\pm$       0.008&       0.906$\pm$       0.007&       0.906$\pm$       0.007&       0.569$\pm$       0.011
\\
   10.0&    1.0&       0.994$\pm$       0.002&       0.996$\pm$       0.001&       0.995$\pm$       0.002&       0.960$\pm$       0.004
\\
   20.0&    1.0&       1&       1&       1&       1
\\
\\
    0.0&    2.0&       0.056$\pm$       0.005&       0.094$\pm$       0.007&       0.092$\pm$       0.006&       0.004$\pm$       0.001
\\
    0.1&    2.0&       0.049$\pm$       0.005&       0.091$\pm$       0.006&       0.088$\pm$       0.006&       0.003$\pm$       0.001
\\
    0.2&    2.0&       0.076$\pm$       0.006&       0.126$\pm$       0.007&       0.120$\pm$       0.007&       0.007$\pm$       0.002
\\
    1.0&    2.0&       0.153$\pm$       0.008&       0.226$\pm$       0.009&       0.214$\pm$       0.009&       0.031$\pm$       0.004
\\
    2.0&    2.0&       0.318$\pm$       0.010&       0.412$\pm$       0.011&       0.394$\pm$       0.011&       0.094$\pm$       0.007
\\
    5.0&    2.0&       0.753$\pm$       0.010&       0.820$\pm$       0.009&       0.803$\pm$       0.009&       0.481$\pm$       0.011
\\
   10.0&    2.0&       0.984$\pm$       0.003&       0.988$\pm$       0.002&       0.987$\pm$       0.003&       0.935$\pm$       0.006
\\
   20.0&    2.0&       1&       1&       1&       1
\\ \hline
\end {tabular}
\end{center}
\newpage{}
\newpage{}
\begin {center}
\noindent {Table   10. Estimated detection probabilities corresponding to uncertainty intervals with
nominal frequentist coverage of 0.90.
Approximate 68 percent uncertainty
due to sampling variability given for cases where estimated coverage probability
is greater than 0 or less than 1.
$\frac{T_{bg}}{T} = 25$.
}
\\
\begin{tabular}{cccccc} \hline
\\
\\
 $\mu_ S$ &  $\mu_B$ & Bayesian     & FC &    RFC &   POE
\\
\\
    0.0&    0.2&       0.018$\pm$       0.003&       0.027$\pm$       0.004&       0.027$\pm$       0.004&       0
\\
    0.1&    0.2&       0.043$\pm$       0.005&       0.067$\pm$       0.006&       0.067$\pm$       0.006&       0
\\
    0.2&    0.2&       0.079$\pm$       0.006&       0.099$\pm$       0.007&       0.099$\pm$       0.007&       0.001$\pm$       0.001
\\
    1.0&    0.2&       0.364$\pm$       0.011&       0.397$\pm$       0.011&       0.397$\pm$       0.011&       0.022$\pm$       0.003
\\
    2.0&    0.2&       0.638$\pm$       0.011&       0.666$\pm$       0.011&       0.666$\pm$       0.011&       0.136$\pm$       0.008
\\
    5.0&    0.2&       0.968$\pm$       0.004&       0.970$\pm$       0.004&       0.970$\pm$       0.004&       0.689$\pm$       0.010
\\
   10.0&    0.2&       1&       1&       1&       0.983$\pm$       0.003
\\
   20.0&    0.2&       1&       1&       1&       1
\\
    0.0&    1.0&       0.043$\pm$       0.005&       0.060$\pm$       0.005&       0.060$\pm$       0.005&       0.001$\pm$       0.001
\\
    0.1&    1.0&       0.055$\pm$       0.005&       0.081$\pm$       0.006&       0.081$\pm$       0.006&       0.002$\pm$       0.001
\\
    0.2&    1.0&       0.076$\pm$       0.006&       0.098$\pm$       0.007&       0.098$\pm$       0.007&       0.004$\pm$       0.001
\\
    1.0&    1.0&       0.209$\pm$       0.009&       0.261$\pm$       0.010&       0.261$\pm$       0.010&       0.022$\pm$       0.003
\\
    2.0&    1.0&       0.417$\pm$       0.011&       0.493$\pm$       0.011&       0.493$\pm$       0.011&       0.097$\pm$       0.007
\\
    5.0&    1.0&       0.887$\pm$       0.007&       0.915$\pm$       0.006&       0.915$\pm$       0.006&       0.582$\pm$       0.011
\\
   10.0&    1.0&       0.997$\pm$       0.001&       0.997$\pm$       0.001&       0.997$\pm$       0.001&       0.972$\pm$       0.004
\\
   20.0&    1.0&       1&       1&       1&       1
\\
    0.0&    2.0&       0.052$\pm$       0.005&       0.071$\pm$       0.006&       0.066$\pm$       0.006&       0.002$\pm$       0.001
\\
    0.1&    2.0&       0.059$\pm$       0.005&       0.079$\pm$       0.006&       0.075$\pm$       0.006&       0.006$\pm$       0.002
\\
    0.2&    2.0&       0.073$\pm$       0.006&       0.092$\pm$       0.006&       0.089$\pm$       0.006&       0.008$\pm$       0.002
\\
    1.0&    2.0&       0.162$\pm$       0.008&       0.203$\pm$       0.009&       0.197$\pm$       0.009&       0.029$\pm$       0.004
\\
    2.0&    2.0&       0.347$\pm$       0.011&       0.404$\pm$       0.011&       0.395$\pm$       0.011&       0.098$\pm$       0.007
\\
    5.0&    2.0&       0.788$\pm$       0.009&       0.823$\pm$       0.009&       0.822$\pm$       0.009&       0.509$\pm$       0.011
\\
   10.0&    2.0&       0.989$\pm$       0.002&       0.991$\pm$       0.002&       0.991$\pm$       0.002&       0.933$\pm$       0.006
\\
   20.0&    2.0&       1&       1&       1&       1
\\ \hline
\end {tabular}
\end{center}

\newpage{}

\begin {center}
\noindent {Table   11. Estimated detection probabilities corresponding to uncertainty intervals with
nominal frequentist coverage of 0.95.
Approximate 68 percent uncertainty
due to sampling variability given for cases where estimated coverage probability
is greater than 0 or less than 1.
$\frac{T_{bg}}{T} = 1$.
}
\\
\begin{tabular}{cccccc} \hline
\\
\\
 $\mu_ S$ &  $\mu_B$ & Bayesian     & FC &    RFC &   POE
\\
\\
    0.0&    0.2&       0&       0.143$\pm$       0.008&       0.143$\pm$       0.008&       0
\\
    0.1&    0.2&       0&       0.202$\pm$       0.009&       0.202$\pm$       0.009&       0
\\
    0.2&    0.2&       0&       0.268$\pm$       0.010&       0.268$\pm$       0.010&       0
\\
    1.0&    0.2&       0.008$\pm$       0.002&       0.580$\pm$       0.011&       0.575$\pm$       0.011&       0.008$\pm$       0.002
\\
    2.0&    0.2&       0.060$\pm$       0.005&       0.759$\pm$       0.010&       0.740$\pm$       0.010&       0.059$\pm$       0.005
\\
    5.0&    0.2&       0.534$\pm$       0.011&       0.951$\pm$       0.005&       0.920$\pm$       0.006&       0.515$\pm$       0.011
\\
   10.0&    0.2&       0.949$\pm$       0.005&       0.998$\pm$       0.001&       0.994$\pm$       0.002&       0.929$\pm$       0.006
\\
   20.0&    0.2&       1&       1&       1     &   1
\\
\\
    0.0&    1.0&       0.001$\pm$       0.001&       0.252$\pm$       0.010&       0.245$\pm$       0.010&       0.002$\pm$       0.001
\\
    0.1&    1.0&       0.003$\pm$       0.001&       0.268$\pm$       0.010&       0.260$\pm$       0.010&       0.003$\pm$       0.001
\\
    0.2&    1.0&       0.004$\pm$       0.001&       0.280$\pm$       0.010&       0.272$\pm$       0.010&       0.004$\pm$       0.001
\\
    1.0&    1.0&       0.017$\pm$       0.003&       0.381$\pm$       0.011&       0.343$\pm$       0.011&       0.015$\pm$       0.003
\\
    2.0&    1.0&       0.082$\pm$       0.006&       0.495$\pm$       0.011&       0.424$\pm$       0.011&       0.072$\pm$       0.006
\\
    5.0&    1.0&       0.468$\pm$       0.011&       0.820$\pm$       0.009&       0.723$\pm$       0.010&       0.379$\pm$       0.011
\\
   10.0&    1.0&       0.917$\pm$       0.006&       0.985$\pm$       0.003&       0.963$\pm$       0.004&       0.844$\pm$       0.008
\\
   20.0&    1.0&       1&       1&       1&       0.999$\pm$       0.001
\\
\\
    0.0&    2.0&       0.007$\pm$       0.002&       0.165$\pm$       0.008&       0.136$\pm$       0.008&       0.006$\pm$       0.002
\\
    0.1&    2.0&       0.013$\pm$       0.003&       0.172$\pm$       0.008&       0.142$\pm$       0.008&       0.012$\pm$       0.002
\\
    0.2&    2.0&       0.011$\pm$       0.002&       0.183$\pm$       0.009&       0.152$\pm$       0.008&       0.009$\pm$       0.002
\\
    1.0&    2.0&       0.033$\pm$       0.004&       0.249$\pm$       0.010&       0.188$\pm$       0.009&       0.025$\pm$       0.003
\\
    2.0&    2.0&       0.100$\pm$       0.007&       0.374$\pm$       0.011&       0.279$\pm$       0.010&       0.066$\pm$       0.006
\\
    5.0&    2.0&       0.405$\pm$       0.011&       0.702$\pm$       0.010&       0.580$\pm$       0.011&       0.285$\pm$       0.010
\\
   10.0&    2.0&       0.837$\pm$       0.008&       0.951$\pm$       0.005&       0.903$\pm$       0.007&       0.742$\pm$       0.010
\\
   20.0&    2.0&       0.997$\pm$       0.001&       0.999$\pm$       0.001&       0.998$\pm$       0.001&       0.993$\pm$       0.002
\\ \hline
\end {tabular}
\end{center}
\newpage{}

\begin {center}
\noindent {Table   12. Estimated detection probabilities corresponding to uncertainty intervals with
nominal frequentist coverage of 0.95.
Approximate 68 percent uncertainty
due to sampling variability given for cases where estimated coverage probability
is greater than 0 or less than 1.
$\frac{T_{bg}}{T} = 5$.
}
\\
\begin{tabular}{cccccc} \hline
\\
\\
 $\mu_ S$ &  $\mu_B$ & Bayesian     & FC &    RFC &   POE
\\
    0.0&    0.2&       0.003$\pm$       0.001&       0.068$\pm$       0.006&       0.068$\pm$       0.006&       0
\\
    0.1&    0.2&       0.020$\pm$       0.003&       0.112$\pm$       0.007&       0.112$\pm$       0.007&       0
\\
    0.2&    0.2&       0.026$\pm$       0.004&       0.140$\pm$       0.008&       0.139$\pm$       0.008&       0
\\
    1.0&    0.2&       0.196$\pm$       0.009&       0.412$\pm$       0.011&       0.410$\pm$       0.011&       0.005$\pm$       0.002
\\
    2.0&    0.2&       0.441$\pm$       0.011&       0.639$\pm$       0.011&       0.635$\pm$       0.011&       0.048$\pm$       0.005
\\
    5.0&    0.2&       0.896$\pm$       0.007&       0.945$\pm$       0.005&       0.943$\pm$       0.005&       0.480$\pm$       0.011
\\
   10.0&    0.2&       0.997$\pm$       0.001&       0.999$\pm$       0.001&       0.998$\pm$       0.001&       0.947$\pm$       0.005
\\
   20.0&    0.2&       1&       1&       1&       1
\\
\\
    0.0&    1.0&       0.017$\pm$       0.003&       0.048$\pm$       0.005&       0.038$\pm$       0.004&       0.001$\pm$       0.001
\\
    0.1&    1.0&       0.024$\pm$       0.003&       0.059$\pm$       0.005&       0.049$\pm$       0.005&       0.001$\pm$       0.001
\\
    0.2&    1.0&       0.029$\pm$       0.004&       0.070$\pm$       0.006&       0.057$\pm$       0.005&       0.001$\pm$       0.001
\\
    1.0&    1.0&       0.104$\pm$       0.007&       0.208$\pm$       0.009&       0.174$\pm$       0.008&       0.008$\pm$       0.002
\\
    2.0&    1.0&       0.283$\pm$       0.010&       0.422$\pm$       0.011&       0.378$\pm$       0.011&       0.047$\pm$       0.005
\\
    5.0&    1.0&       0.782$\pm$       0.009&       0.865$\pm$       0.008&       0.846$\pm$       0.008&       0.398$\pm$       0.011
\\
   10.0&    1.0&       0.985$\pm$       0.003&       0.994$\pm$       0.002&       0.991$\pm$       0.002&       0.916$\pm$       0.006
\\
   20.0&    1.0&       1&       1&       1&       1
\\
\\
    0.0&    2.0&       0.020$\pm$       0.003&       0.061$\pm$       0.005&       0.043$\pm$       0.005&       0.001$\pm$       0.001
\\
    0.1&    2.0&       0.026$\pm$       0.004&       0.051$\pm$       0.005&       0.039$\pm$       0.004&       0.001$\pm$       0.001
\\
    0.2&    2.0&       0.034$\pm$       0.004&       0.075$\pm$       0.006&       0.061$\pm$       0.005&       0.001$\pm$       0.001
\\
    1.0&    2.0&       0.084$\pm$       0.006&       0.145$\pm$       0.008&       0.125$\pm$       0.007&       0.010$\pm$       0.002
\\
    2.0&    2.0&       0.208$\pm$       0.009&       0.305$\pm$       0.010&       0.264$\pm$       0.010&       0.039$\pm$       0.004
\\
    5.0&    2.0&       0.635$\pm$       0.011&       0.743$\pm$       0.010&       0.698$\pm$       0.010&       0.324$\pm$       0.010
\\
   10.0&    2.0&       0.970$\pm$       0.004&       0.982$\pm$       0.003&       0.979$\pm$       0.003&       0.871$\pm$       0.008
\\
   20.0&    2.0&       1&       1&       1&       0.999$\pm$       0.001
\\ \hline
\end {tabular}
\end{center}
\newpage{}
\begin {center}
\noindent {Table   13. Estimated detection probabilities corresponding to uncertainty intervals with
nominal frequentist coverage of 0.95.
Approximate 68 percent uncertainty
due to sampling variability given for cases where estimated coverage probability
is greater than 0 or less than 1.
$\frac{T_{bg}}{T} = 25$.
}
\\
\begin{tabular}{cccccc} \hline
\\
\\
 $\mu_ S$ &  $\mu_B$ & Bayesian     & FC &    RFC &   POE
\\
\\
    0.0&    0.2&       0.010$\pm$       0.002&       0.017$\pm$       0.003&       0.017$\pm$       0.003&      0 
\\
    0.1&    0.2&       0.024$\pm$       0.003&       0.042$\pm$       0.004&       0.042$\pm$       0.004&       0
\\
    0.2&    0.2&       0.058$\pm$       0.005&       0.074$\pm$       0.006&       0.074$\pm$       0.006&       0
\\
    1.0&    0.2&       0.304$\pm$       0.010&       0.352$\pm$       0.011&       0.352$\pm$       0.011&       0.002$\pm$       0.001
\\
    2.0&    0.2&       0.573$\pm$       0.011&       0.626$\pm$       0.011&       0.626$\pm$       0.011&       0.027$\pm$       0.004
\\
    5.0&    0.2&       0.952$\pm$       0.005&       0.966$\pm$       0.004&       0.966$\pm$       0.004&       0.429$\pm$       0.011
\\
   10.0&    0.2&       1&       1&       1&       0.946$\pm$       0.005
\\
   20.0&    0.2&       1&       1&       1&       1
\\
\\
    0.0&    1.0&       0.019$\pm$       0.003&       0.029$\pm$       0.004&       0.028$\pm$       0.004&       0
\\
    0.1&    1.0&       0.030$\pm$       0.004&       0.042$\pm$       0.004&       0.038$\pm$       0.004&       0
\\
    0.2&    1.0&       0.037$\pm$       0.004&       0.059$\pm$       0.005&       0.052$\pm$       0.005&       0.001$\pm$       0.001
\\
    1.0&    1.0&       0.131$\pm$       0.008&       0.160$\pm$       0.008&       0.152$\pm$       0.008&       0.003$\pm$       0.001
\\
    2.0&    1.0&       0.333$\pm$       0.011&       0.372$\pm$       0.011&       0.363$\pm$       0.011&       0.037$\pm$       0.004
\\
    5.0&    1.0&       0.838$\pm$       0.008&       0.863$\pm$       0.008&       0.855$\pm$       0.008&       0.393$\pm$       0.011
\\
   10.0&    1.0&       0.996$\pm$       0.001&       0.996$\pm$       0.001&       0.996$\pm$       0.001&       0.928$\pm$       0.006
\\
   20.0&    1.0&       1&       1&       1&       1
\\
\\
    0.0&    2.0&       0.022$\pm$       0.003&       0.036$\pm$       0.004&       0.031$\pm$       0.004&       0.001$\pm$       0.001
\\
    0.1&    2.0&       0.033$\pm$       0.004&       0.045$\pm$       0.005&       0.042$\pm$       0.004&       0.001$\pm$       0.001
\\
    0.2&    2.0&       0.036$\pm$       0.004&       0.053$\pm$       0.005&       0.051$\pm$       0.005&       0.001$\pm$       0.001
\\
    1.0&    2.0&       0.105$\pm$       0.007&       0.132$\pm$       0.008&       0.126$\pm$       0.007&       0.010$\pm$       0.002
\\
    2.0&    2.0&       0.243$\pm$       0.010&       0.295$\pm$       0.010&       0.286$\pm$       0.010&       0.035$\pm$       0.004
\\
    5.0&    2.0&       0.703$\pm$       0.010&       0.738$\pm$       0.010&       0.731$\pm$       0.010&       0.340$\pm$       0.011
\\
   10.0&    2.0&       0.976$\pm$       0.003&       0.984$\pm$       0.003&       0.983$\pm$       0.003&       0.855$\pm$       0.008
\\
   20.0&    2.0&       1&       1&       1&       1
\\ \hline
\end {tabular}
\end{center}
\newpage{}
\begin {center}
\noindent {Table   14. 
Root-mean-square deviation between observed and
nominal frequentist coverage averaged over all
values of $\mu_S \le $ 10.
Nominal frequentist coverage is 0.90.
}
\\
\begin{tabular}{cccccc} \hline
\\
\\
$\frac{T_{bg}}{T}$ &  $\mu_B$ & Bayesian     & FC &    RFC &   POE
\\
\\
  1& 0.2&    0.084&    0.052&    0.053&    0.087
\\
  1& 1.0&    0.075&    0.101&    0.090&    0.058
\\
  1& 2.0&    0.062&    0.077&    0.045&    0.044
\\
\\
  2& 0.2&    0.080&    0.044&    0.049&    0.101
\\
  2& 1.0&    0.064&    0.030&    0.018&    0.054
\\
  2& 2.0&    0.055&    0.031&    0.017&    0.040
\\
\\
  5& 0.2&    0.067&    0.039&    0.039&    0.107
\\
  5& 1.0&    0.055&    0.023&    0.028&    0.044
\\
  5& 2.0&    0.046&    0.013&    0.021&    0.039
\\
\\
 10& 0.2&    0.065&    0.038&    0.042&    0.105
\\
 10& 1.0&    0.054&    0.026&    0.031&    0.077
\\
 10& 2.0&    0.049&    0.022&    0.031&    0.038
\\
\\
 25& 0.2&    0.061&    0.055&    0.055&    0.101
\\
 25& 1.0&    0.050&    0.033&    0.034&    0.122
\\
 25& 2.0&    0.044&    0.024&    0.026&    0.039
\\ \hline
\end {tabular}
\end{center}

\newpage{}
\newpage{}
\begin {center}
\noindent {Table   15. 
Root-mean-square deviation between observed and
nominal frequentist coverage averaged over all
values of $\mu_S \le $ 10.
Nominal frequentist coverage is 0.95.
}
\\
\begin{tabular}{cccccc} \hline
\\
\\
$\frac{T_{bg}}{T}$ &  $\mu_B$ & Bayesian     & FC &    RFC &   POE
\\
\\
  1& 0.2&    0.042&    0.040&    0.040&    0.084
\\
  1& 1.0&    0.040&    0.088&    0.084&    0.031
\\
  1& 2.0&    0.035&    0.066&    0.046&    0.027
\\
\\
  2& 0.2&    0.040&    0.031&    0.031&    0.105
\\
  2& 1.0&    0.032&    0.030&    0.025&    0.047
\\
  2& 2.0&    0.028&    0.024&    0.014&    0.020
\\
\\
  5& 0.2&    0.035&    0.019&    0.019&    0.110
\\
  5& 1.0&    0.029&    0.011&    0.014&    0.043
\\
  5& 2.0&    0.026&    0.007&    0.011&    0.035
\\
\\
 10& 0.2&    0.035&    0.020&    0.020&    0.107
\\
 10& 1.0&    0.027&    0.016&    0.019&    0.071
\\
 10& 2.0&    0.026&    0.014&    0.017&    0.040
\\
\\
 25& 0.2&    0.031&    0.024&    0.024&    0.103
\\
 25& 1.0&    0.027&    0.021&    0.022&    0.138
\\
 25& 2.0&    0.024&    0.011&    0.014&    0.043
\\ \hline
\end {tabular}
\end{center}
\section{References}

\noindent{}
   [1]Mandelkern M. 2002
   ``Setting confidence intervals for bounded parameters"
   Stat. Sci.
   Vol. 17 No. 2
   p. 149

\noindent{}
   [2]Abdurashitov J.N.  et al.
``Measurement of the solar neutrino capture rate by SAGE
and implications for neutrino oscillations in vacuum"
   1999
   Phys. Rev. Lett.
   83
   p. 4686

\noindent{}
[3]   Ahmad  Q.R., et al.
``Direct Evidence for Neutrino Flavor
Transformation from neutral-current interactions
in the Sudbury Neutrino Observatory"
2002
   Phys. Rev. Lett.
   89
   011301

\noindent{}
[4]   Altman M and et al.
``GNO Solar neutrino observations:
results for GNO I"
2000
   Phys. Lett. B
   490
   p. 16

\noindent{}
   [5] Athanassopoulos C et al.
``Results on $\nu_{\mu} \rightarrow \mu_{e}$
neutrino oscillations from the LSND Experiment"
   1998
   Phys. Rev. Lett.
   81
   p. 1774

\noindent{}
   [6] Cleveland B.T. et al.
``Measurements of the solar electron neutrino
flux with the Homestake chlorine detector"
   1998
   Astrophys. J.
   496
   p. 505

\noindent{}
[7] Alner G.J. et al.
``Status of the Zeplin II Experiment"
2005
New Astronomy Reviews
49
p. 245

\noindent{}
[8] Abrams D.et al.
``Exclusion limits on the WIMP-nucleon cross section from the Cryogenic Dark Matter Search"
2002
Phys. Rev. D.
66
122003

\noindent{}
[9] Conrad J., Scargel J., Ylilen, T.
``Statistical analysis of detection of, and upper limits on, dark matter
lines"
2007
Thd First GLAST Symposium,
edited by S. Ritz, P. Michelson, and C. Meegan,
5-8 February 2007 Stanford University
AIP Conference Proceedings 2007 Vol. 921, Issue 1 p. 586.

\noindent{}
[10]
Richter S et al.
``Isotopic ``fingerprints'' for natural uranium ore samples"
1999
Int. J. Mass Spectrom.
Vol. 193, issue 1,
p. 9

\noindent{}
[11] Hotchkis M.A.C. et al.
``Measurement of ${}^{236}$U in environmental media"
2000
Nucl. Instrum. Meth. Phys. Res. B
Vol. 172,
Issue 1-4,
p. 659

\noindent{}
[12] Currie L.A.
``The measurement of environmental levels of rare gas nuclides and the
treatment of very low-level counting data"
1972
IEEE Trans. Nucl. Sci.
Vol. 19
Issue 1
p. 119

\noindent{}
[13] Currie L.A.
``Detection and quantification limits: basic concepts, international harmonization, and 
outstanding (``low level") issues"
2004
App. Rad. and Isotopes
Vol. 61
Issues 2-3
p. 145

\noindent{}
[14] Currie L.A.
``On the detection of rare, and moderately rare, nuclear events"
J. of Radio. and Nucl. Chem.
2008
Vol. 276
No. 2
p. 285

\noindent{}
[15] DeGeer L.E.
``Currie detection limits in gamma-ray spectroscopy"
2004
App. Rad. and Isotopes
Vol. 61
Issues 2-3
p. 151

\noindent{}
[16]
Liu H.K and  Ehara K.
1996
``Background corrected statistical confidence intervals for particle contamination levels"
Proceedings of the 13th International Symposium on Contamination Control
p. 478

\noindent{}
[17] Feldman G.J. 
and
Cousins R.D.
1998
``Unified approach to the classical statistical analysis of small signals"
Phys. Rev. D
Vol. 57
Issue 7
p. 3873

\noindent{}
[18] Neyman J.
 "Outline of a theory of statistical estimation based on the classical theory of probability"
1937
Philos. Trans. Roy. Soc. London
Ser A
236 
p. 333

\noindent{}
[19]
Conrad J. Botner O. Hallgren A. Pérez de los Heros C
``Including systematic uncertainties in confidence interval construction for Poisson statistics"
2003
Phys. Rev. D
67
012002

\noindent{}
[20] 
Efron B. and Tibshirani R.J.
An Introduction to the Bootstrap
Monographs on Statistics and Applied Probability Series
Vol. 57
Chapman and Hall 
New York
1994

\noindent{}
[21]
Box G.E. Tiao G.C.
1992
Bayesian Inference in Statistical Analysis, New York: Wiley

\noindent{}
[22]  Gelman A. Carlin J.B. Stern H.S. Rubin D.B.
2004
Bayesian Data Analysis, Second Edition , Boca Raton: Chapman and Hall/CRC

\noindent{}
[23]
Loredo T.J.
1992
The Promise of Bayesian Inference For Astrophysics, Statistical Challenges in Modern Astronomy, E.D.Feigelson and G.J.Babu (eds.),
Springer-Verlag

\noindent{}
[24] Roe B. P. Woodroofe M. B. 
``Setting confidence belts"
2000
Phys. Rev. D.
63
013009

\noindent{}
[25] Bayarri M.J. Berger J.
2004
``The interplay of Bayesian and Frequentist Analysis"
Statistical Science
19 1
p. 58

\noindent{}
[26] Bernardo J.M.
2006
A Bayesian Mathematical Statistical Primer, Proc. 7th Int. Conf. on Teach Statistics 3I (A.Rossman and B. Chance ,eds)"
Amsterdam: IASE, ISI

\noindent{}
[27]
Cox D.R. and Lewis P.A.W.
1966
The Statistical Analysis of Series of Events,
Metheun

\noindent{}
[28] Perlman M.D.
and
Wu L.
``The emperor's new tests"
1999
Statist. Sci.
14
p.549

\noindent{}
[29] Giunti C.
``New ordering principle for the classical
statistical analysis of Poisson processes
with background"
1999
Phys. Rev. D.
59
053001

\noindent{}
[30] Roe B.P. Woodroofe M.B. 
``Improved probability method for estimating signal in the presence of background,"
1999
Phys. Rev. D.
60
053009

\noindent{}
[31] Hill G.C.  
``Comment on ``Including systematic uncertainties in confidence interval construction for Poisson
statistics"
2003
Phys. Rev. D.
67
118101

\newpage{}

\begin{figure}
\begin{center}
\includegraphics[height = 6.0in]{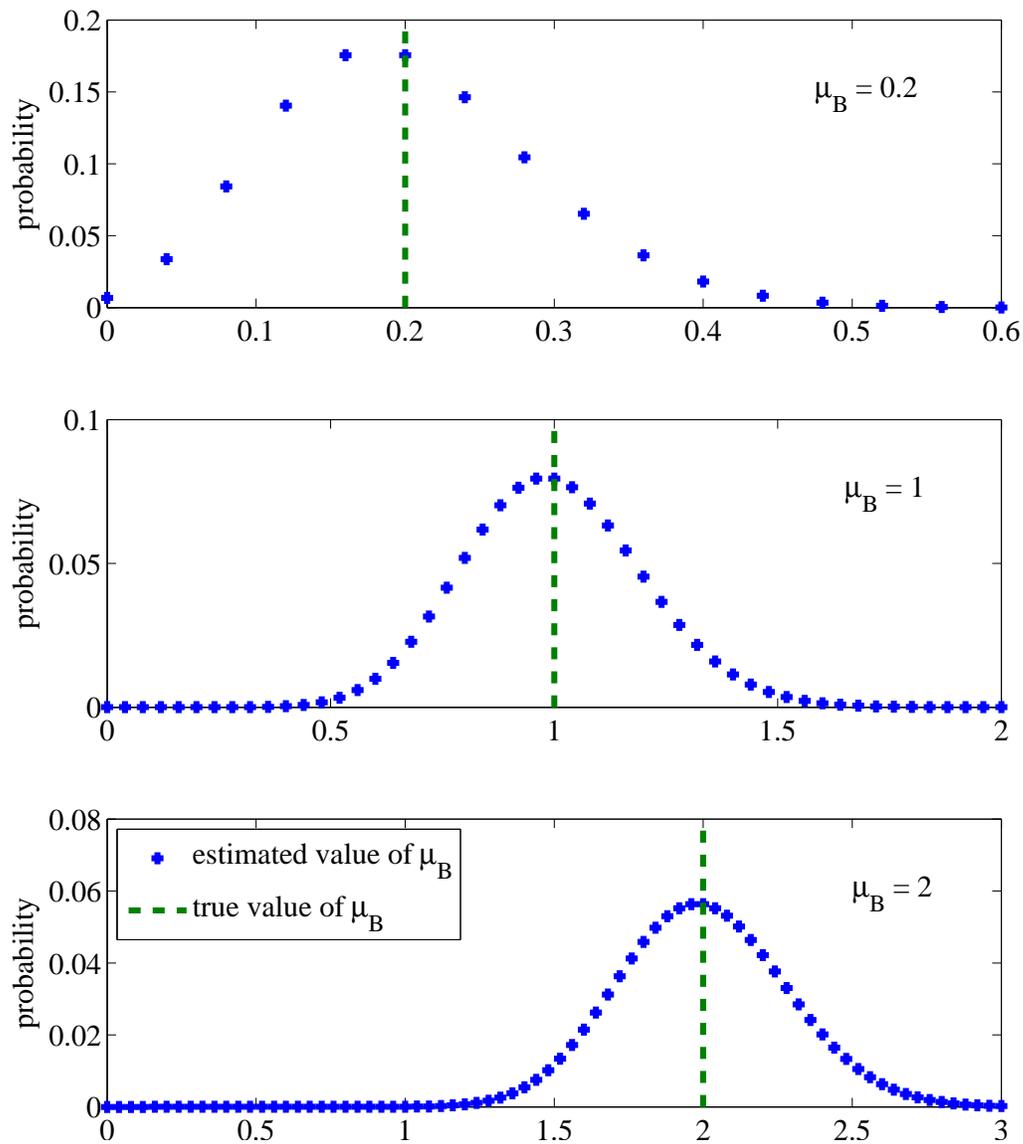}
\caption{
\label{fig:fig1}
Probability density functions for estimated background
for case where $T_{bg}/T = $ 25.
For $\mu_B=$ 0.2, 1, and 2, the standard deviations of 
the estimated background
are 0.09, 0.20 and 0.28 respectively.
The corresponding fractional
standard deviations of the estimated background are
45 $\%$, 20 $\%$, and
14 $\%$ respectively.
}
\end{center}
\end{figure}

\begin{figure}
\begin{center}
\includegraphics[height = 6.0in]{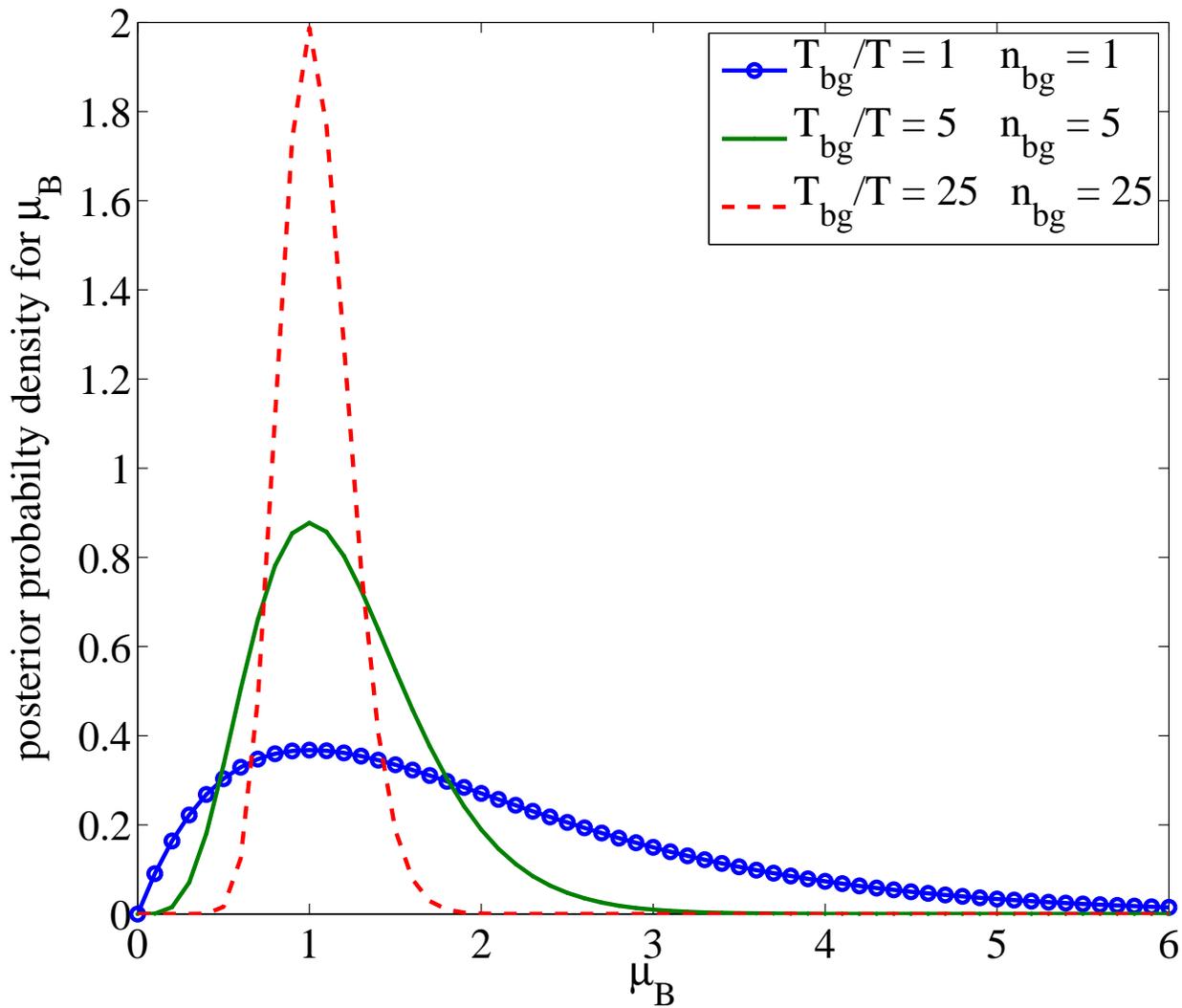}
\caption{
\label{fig:fig1}
Posterior probability density functions
for the background parameter $\mu_B$
given the observed value $n_{bg}$.
}
\end{center}
\end{figure}

\begin{figure}
\begin{center}
\includegraphics[height = 8.0in]{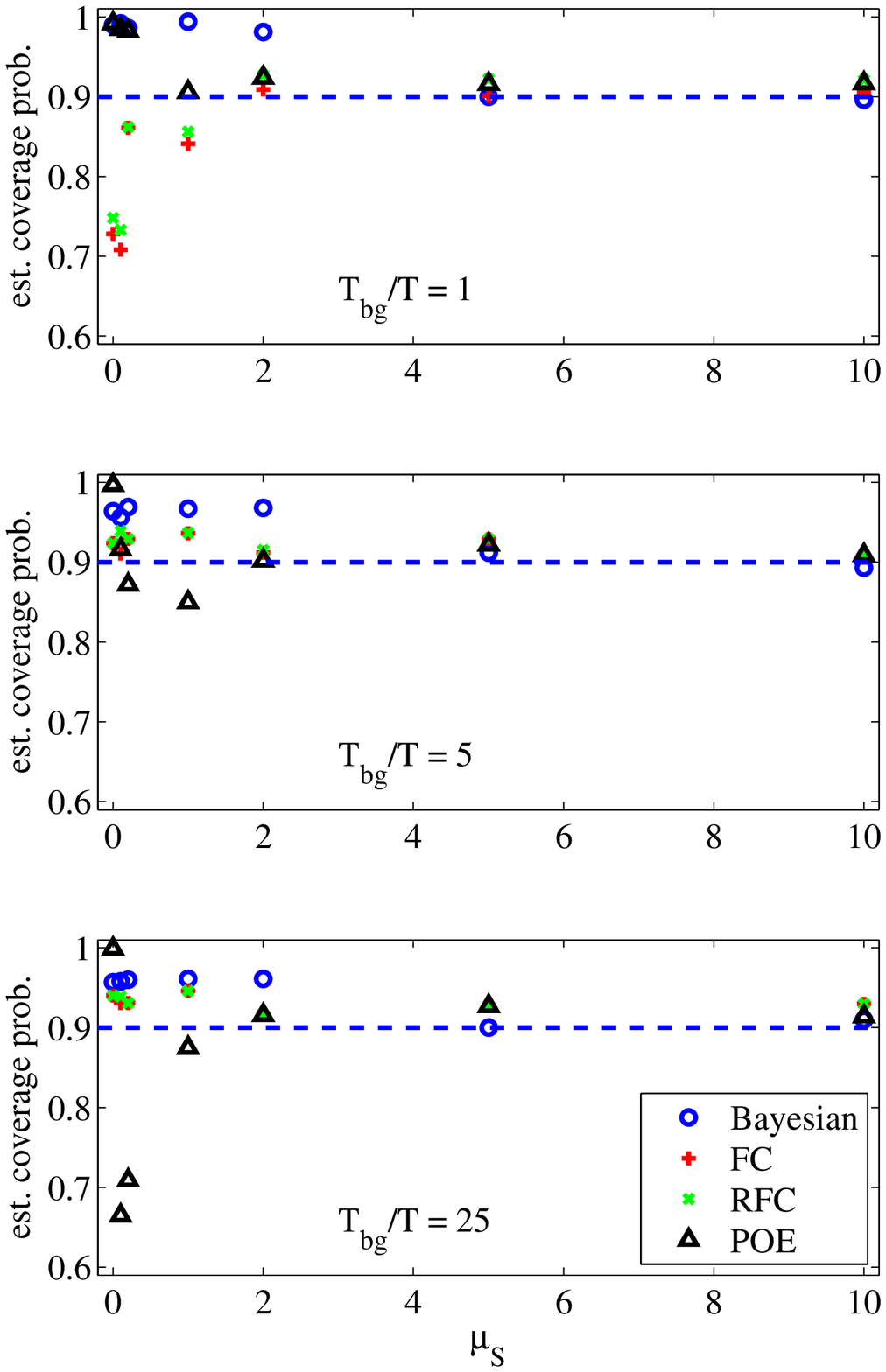}
\caption{
\label{fig:fig3}
Estimated coverage probabilities
for case where 
$\mu_B = 1$ for intervals with
nominal frequentist coverage probability of 0.90.
}
\end{center}
\end{figure}

\begin{figure}
\begin{center}
\includegraphics[height = 8.0in]{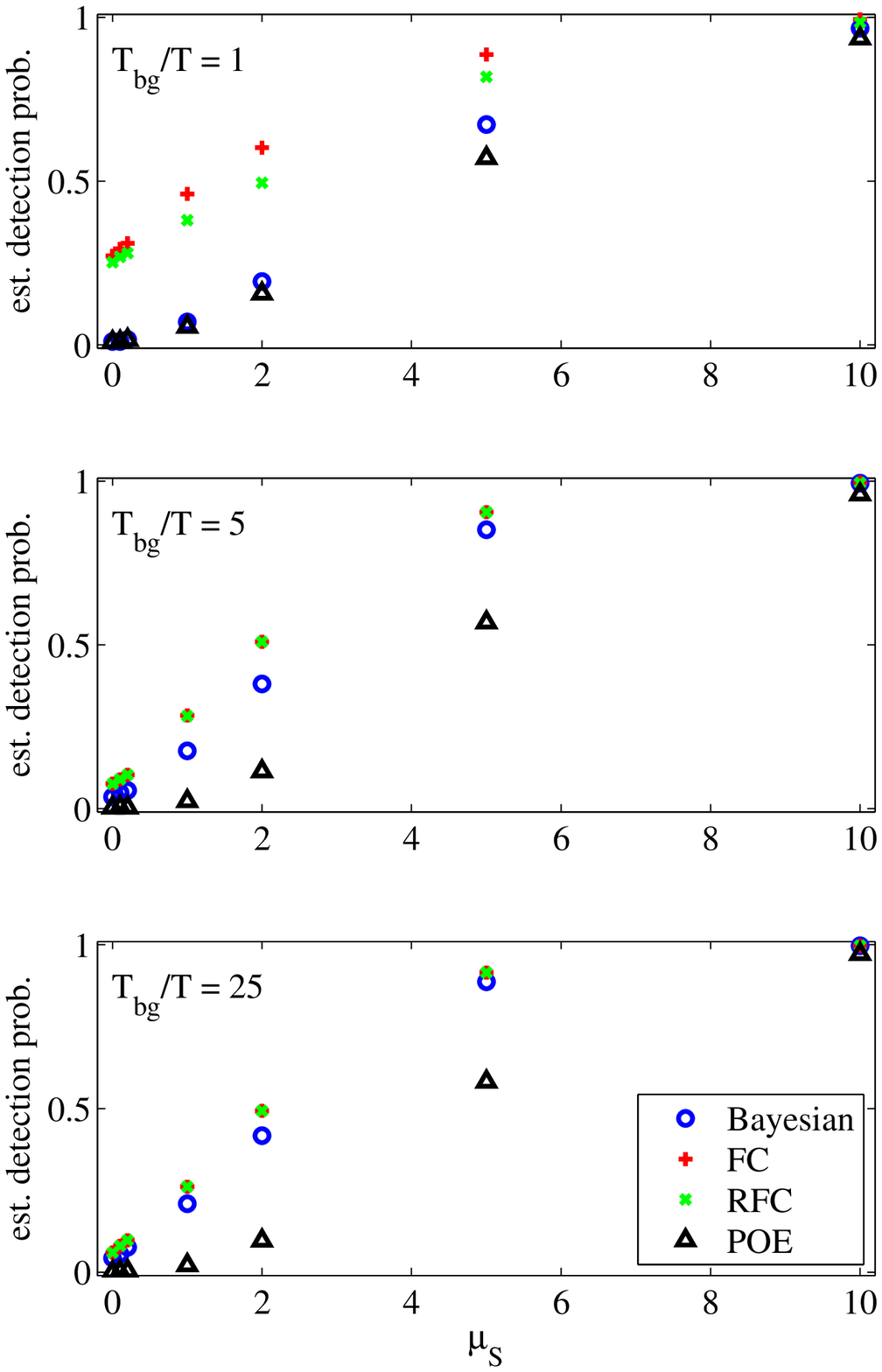}
\caption{
\label{fig:fig3}
Estimated detection probabilities
for case where 
$\mu_B = 1$ for intervals with
nominal frequentist coverage probability of 0.90.
}
\end{center}
\end{figure}

\begin{figure}
\begin{center}
\includegraphics[height = 8.0in]{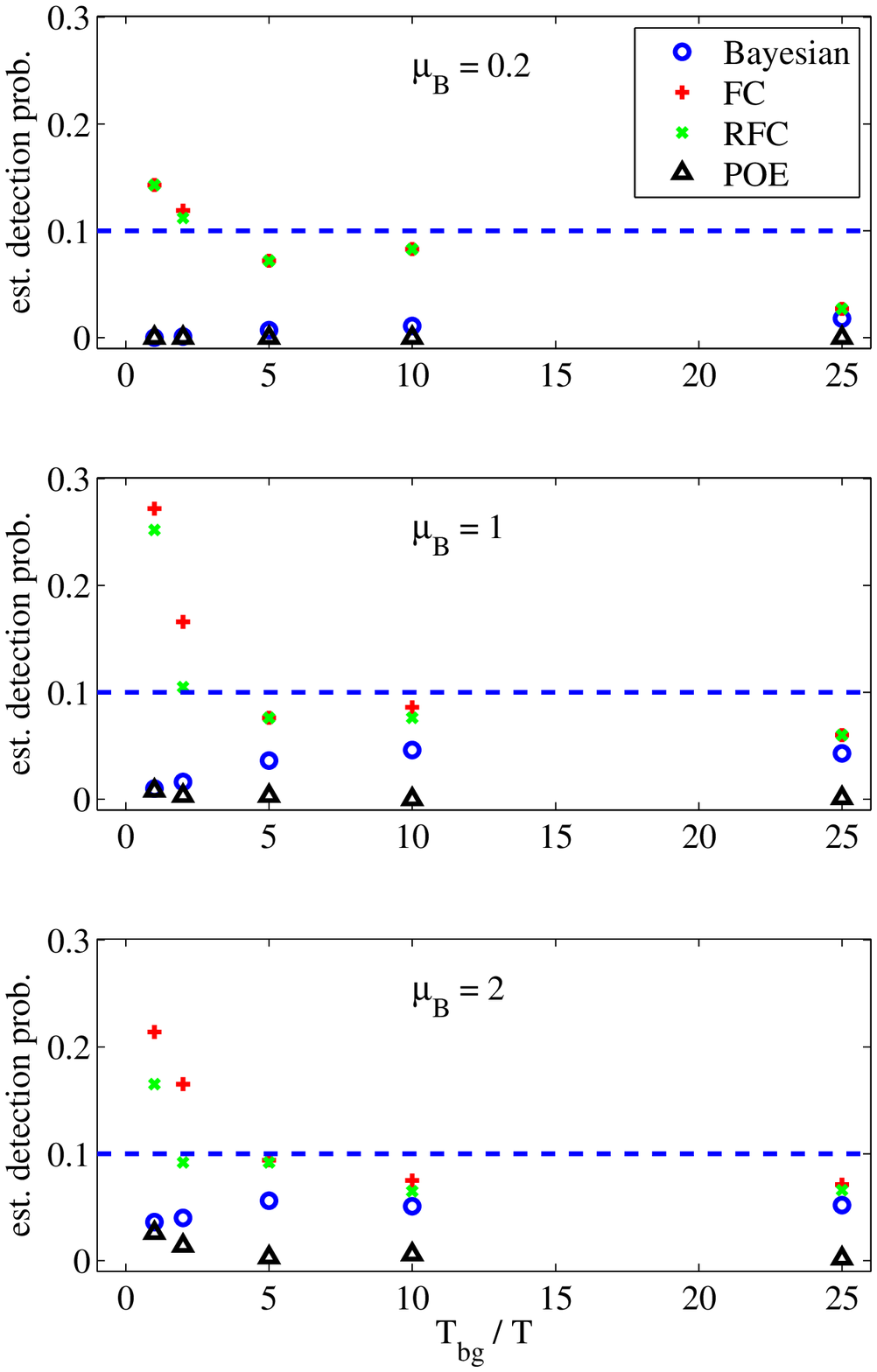}
\caption{
\label{fig:fig3}
Estimated detection probabilities
for case where there is no signal
($\mu_S = 0$) associated with intervals with
nominal frequentist coverage probability of 0.90.
}
\end{center}
\end{figure}

\begin{figure}
\begin{center}
\includegraphics[height = 8.0in]{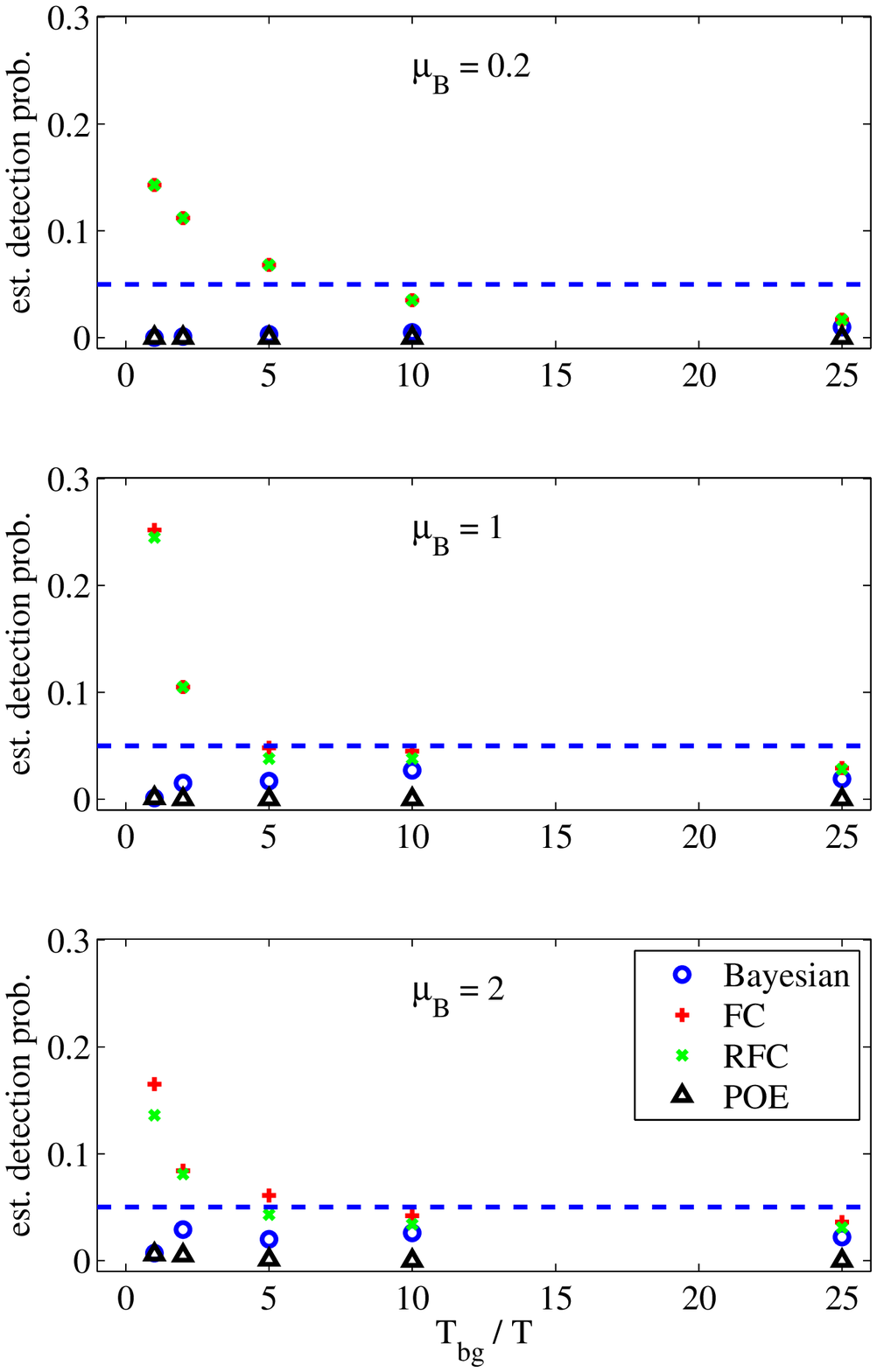}
\caption{
\label{fig:fig3}
Estimated detection probabilities
for case where there is no signal
($\mu_S = 0$) associated with intervals with
nominal frequentist coverage probability of 0.95.
}
\end{center}
\end{figure}

\begin{figure}
\begin{center}
\includegraphics[height = 8.5in]{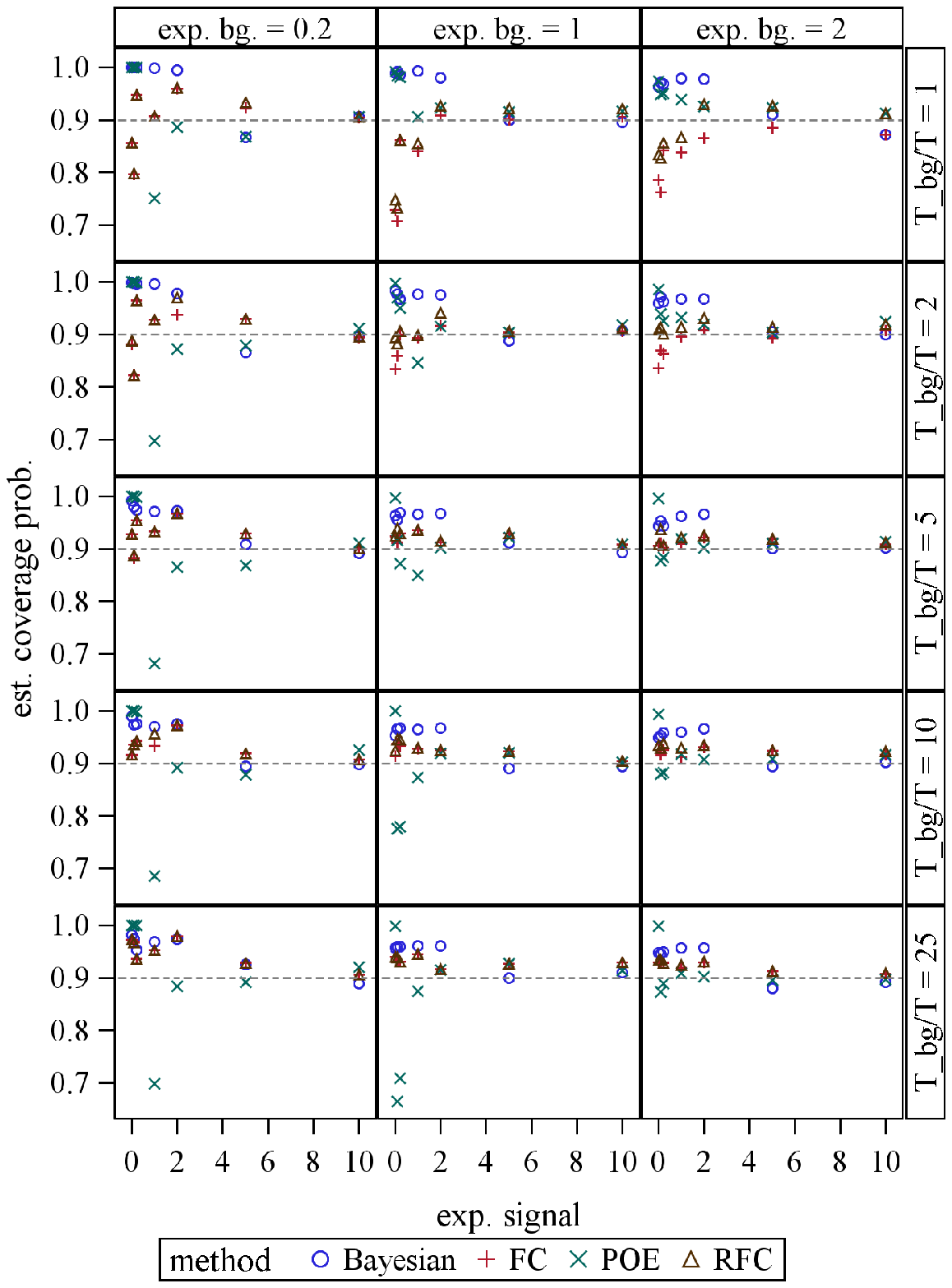}
\caption{
\label{fig:fig2}
Estimated coverage probabilities
corresponding to intervals 
with target coverage of 0.90.
In the plots, we show results for
$\mu_B = 0.2,1,2$ and
$T_{bg}/T  = 1,2,5,10,25$.
}
\end{center}
\end{figure}

\begin{figure}
\begin{center}
\includegraphics[height = 8.5in]{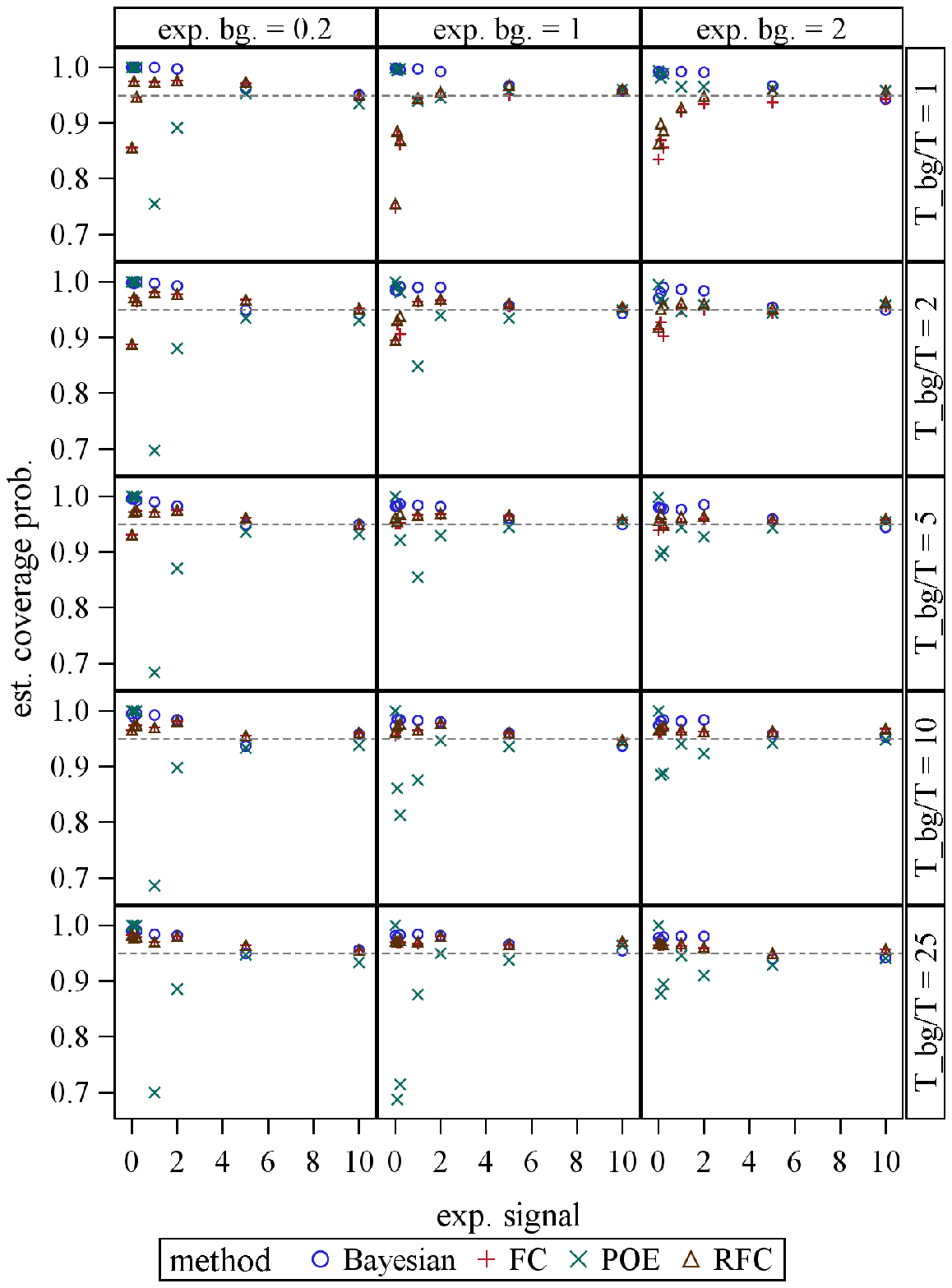}
\caption{
Estimated coverage probabilities
corresponding to intervals 
with target coverage of 0.95.
In the plots, we show results for
$\mu_B = 0.2,1,2$ and
$T_{bg}/T  = 1,2,5,10,25$.
\label{fig:fig2}
}
\end{center}
\end{figure}

\begin{figure}
\begin{center}
\includegraphics[height = 8.5in]{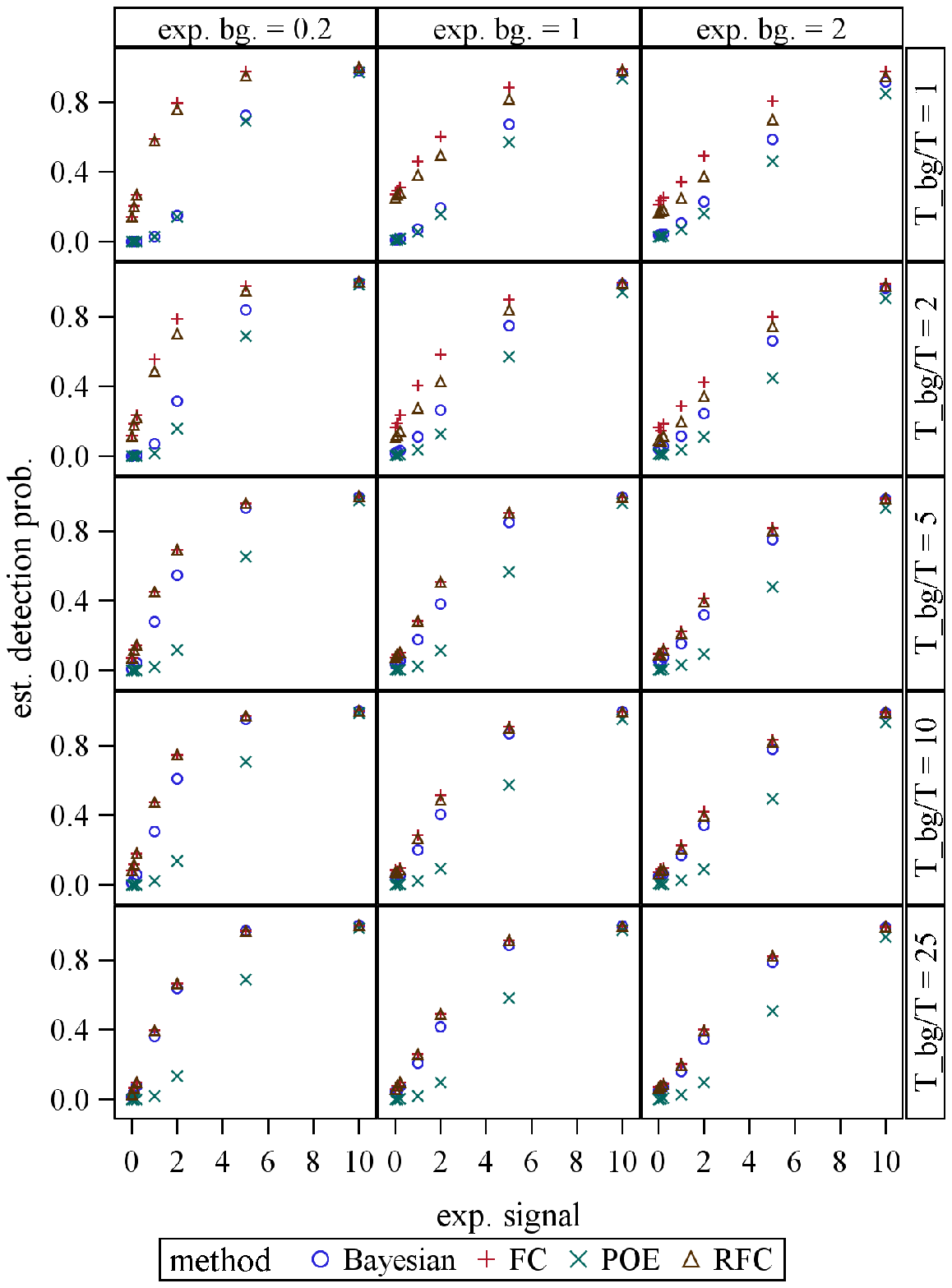}
\caption{
\label{fig:fig2}
Estimated detection probabilities
corresponding to intervals 
with target coverage of 0.90.
In the plots, we show results for
$\mu_B = 0.2,1,2$ and
$T_{bg}/T  = 1,2,5,10,25$.
}
\end{center}
\end{figure}

\begin{figure}
\begin{center}
\includegraphics[height = 8.5in]{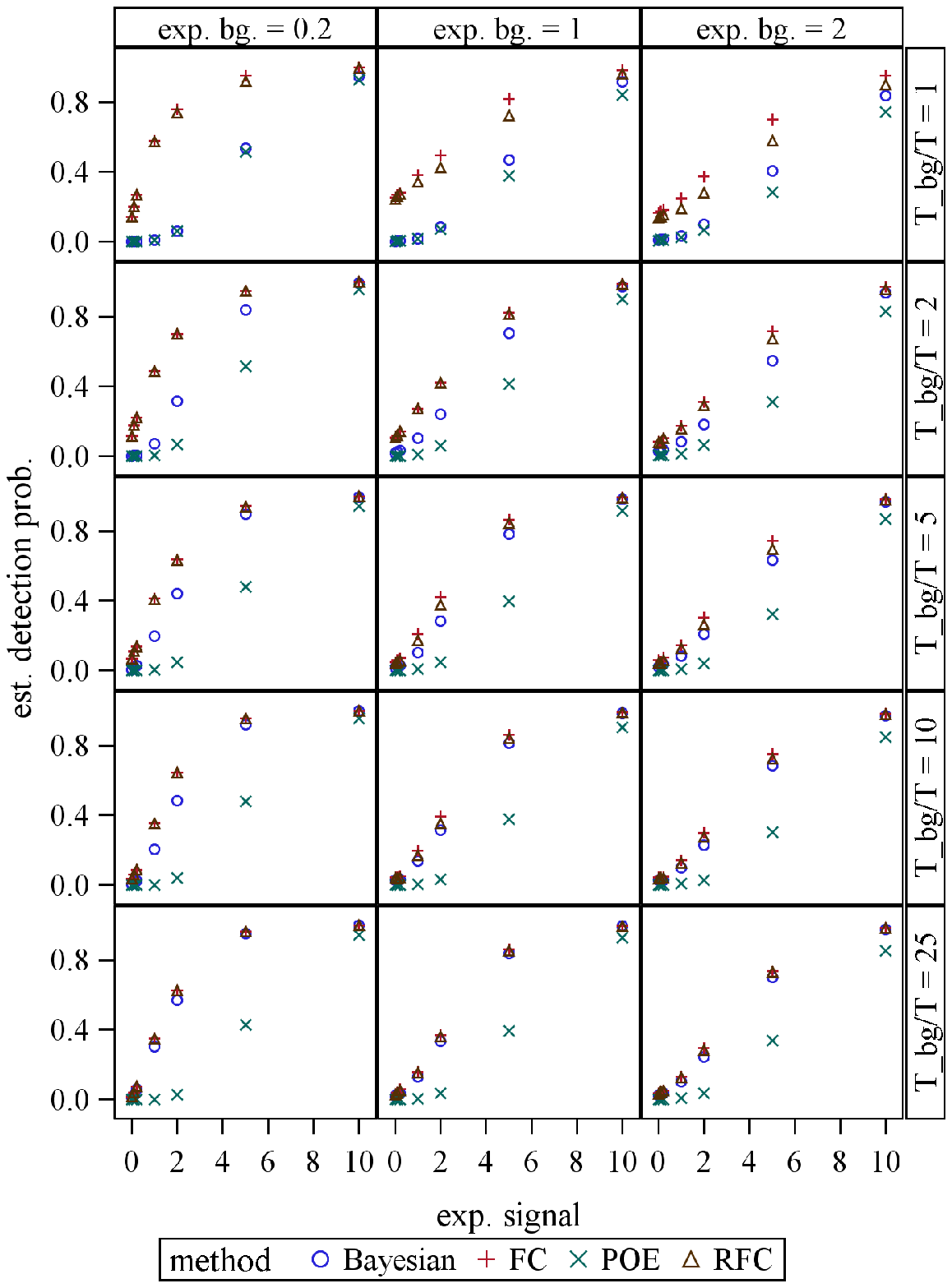}
\caption{
\label{fig:fig3}
Estimated detection probabilities
corresponding to intervals 
with target coverage of 0.95.
In the plots, we show results for
$\mu_B = 0.2,1,2$ and
$T_{bg}/T  = 1,2,5,10,25$.
}
\end{center}
\end{figure}
\end{document}